\documentclass[sigconf, screen]{acmart}

\AtBeginDocument{%
  \providecommand\BibTeX{{%
    \normalfont B\kern-0.5em{\scshape i\kern-0.25em b}\kern-0.8em\TeX}}}

\copyrightyear{2025}
\acmYear{2025}
\setcopyright{cc}
\setcctype{by}
\acmConference[CHI '25]{CHI Conference on Human Factors in Computing Systems}{April 26-May 1, 2025}{Yokohama, Japan}
\acmBooktitle{CHI Conference on Human Factors in Computing Systems (CHI '25), April 26-May 1, 2025, Yokohama, Japan}\acmDOI{10.1145/3706598.3713514}
\acmISBN{979-8-4007-1394-1/25/04}

\usepackage{subcaption} 
\usepackage{url}
\usepackage{verbatim}
\makeatletter
\g@addto@macro{\UrlBreaks}{\UrlOrds}
\makeatother
\usepackage{multirow} 
\usepackage{color} 
\usepackage{enumitem} 

\usepackage{longtable}
\usepackage{xspace}
\usepackage{tabularx}

\usepackage{arydshln}
\usepackage{booktabs}

\usepackage{amsmath}
\DeclareMathOperator*{\argmax}{arg\,max}

\newcommand\itema{\item[\textbf{\textcolor{blue}{RQ1}}]}
\newcommand\itemb{\item[\textbf{\textcolor{blue}{RQ2}}]}
\newcommand\itemc{\item[\textbf{\textcolor{blue}{RQ3}}]}

\newcommand\itemas{\item[\textbf{\textcolor{brown}{C1}}]}
\newcommand\itembs{\item[\textbf{\textcolor{brown}{C2}}]}
\newcommand\itemcs{\item[\textbf{\textcolor{brown}{C3}}]}
\newcommand\itemds{\item[\textbf{\textcolor{brown}{C4}}]}
\newcommand\itemes{\item[\textbf{\textcolor{brown}{C5}}]}
\newcommand\itemfs{\item[\textbf{\textcolor{brown}{C6}}]}

\newcommand\itemha{\item[\textbf{\textcolor{teal}{H1}}]}
\newcommand\itemhb{\item[\textbf{\textcolor{teal}{H2}}]}

\newcommand{\m}{\textit{M=}}

\newcommand{\sd}{\textit{SD=}}

\newcommand{\F}[3]{$F({#1},{#2})={#3}$}
\newcommand{\p}{\textit{p=}}
\newcommand{\padj}{\textit{adj. p=}}
\newcommand{\padjminor}{\textit{adj. p$<$}}

\newcommand{\pminor}{\textit{p$<$}}

\newcommand{\chisq}{$\chi^2$}

\newcommand{\opticar}{\textsc{OptiCarVis}\xspace}

\newcommand{\aoi}{\textit{AOI}\xspace}
\newcommand{\GroupID}{\textit{visualization condition}\xspace}
\newcommand{\optimization}{\textit{visualization condition}\xspace}

\newcommand{\trust}{\textit{trust}\xspace}
\newcommand{\predictability}{\textit{predictability}\xspace}

\newcommand{\perceivedSafetyScore}{\textit{perceived safety}\xspace}

\newcommand{\MentalLoad}{\textit{cognitive load}\xspace}

\begin{document}

\title[\opticar: Bayesian Optimization of Automated Vehicle Functionality Visualizations]{\opticar: Improving Automated Vehicle Functionality Visualizations Using Bayesian Optimization to Enhance User Experience}

\author{Pascal Jansen}
\authornote{Both authors contributed equally to this research.}
\email{pascal.jansen@uni-ulm.de}
\orcid{0000-0002-9335-5462}
\affiliation{%
  \institution{Institute of Media Informatics, Ulm University}
  \city{Ulm}
  \country{Germany}
}

\author{Mark Colley}
\authornotemark[1]
\email{mark.colley@uni-ulm.de}
\orcid{0000-0001-5207-5029}
\affiliation{%
  \institution{Institute of Media Informatics, Ulm University}
  \city{Ulm}
  \country{Germany}
}
\affiliation{%
  \institution{Cornell Tech}
  \streetaddress{2 W Loop Road}
  \city{New York}
  \country{U.S.}
}

\author{Svenja Krauß}
\email{svenja.krauss@uni-ulm.de}
\orcid{0009-0002-4047-0130}
\affiliation{%
  \institution{Institute of Media Informatics, Ulm University}
  \city{Ulm}
  \country{Germany}
}

\author{Daniel Hirschle}
\email{daniel.hirschle@uni-ulm.de}
\orcid{0009-0004-9483-4328}
\affiliation{%
  \institution{Institute of Media Informatics, Ulm University}
  \city{Ulm}
  \country{Germany}
}

\author{Enrico Rukzio}
\email{enrico.rukzio@uni-ulm.de}
\orcid{0000-0002-4213-2226}
\affiliation{%
  \institution{Institute of Media Informatics, Ulm University}
  \city{Ulm}
  \country{Germany}
}

\renewcommand{\shortauthors}{Jansen \& Colley et al.}

\begin{abstract}
Automated vehicle (AV) acceptance relies on their understanding via feedback.
While visualizations aim to enhance user understanding of AV's detection, prediction, and planning functionalities, establishing an optimal design is challenging. Traditional "one-size-fits-all" designs might be unsuitable, stemming from resource-intensive empirical evaluations.
This paper introduces \opticar, a set of Human-in-the-Loop (HITL) approaches using Multi-Objective Bayesian Optimization (MOBO) to optimize AV feedback visualizations. We compare conditions using eight expert and user-customized designs for a \textit{Warm-Start} HITL MOBO.
An online study (N=117) demonstrates \opticar's efficacy in significantly improving trust, acceptance, perceived safety, and predictability without increasing cognitive load.
\opticar facilitates a comprehensive design space exploration, enhancing in-vehicle interfaces for optimal passenger experiences and broader applicability.
\end{abstract}

\ccsdesc[500]{Human-centered computing~Systems and tools for interaction design}
\ccsdesc[300]{Human-centered computing~Empirical studies in visualization}
\ccsdesc[500]{Human-centered computing~Empirical studies in HCI}

\keywords{automated vehicles, user study, bayesian optimization, multi objective}

\begin{teaserfigure}
    \includegraphics[width=\linewidth]{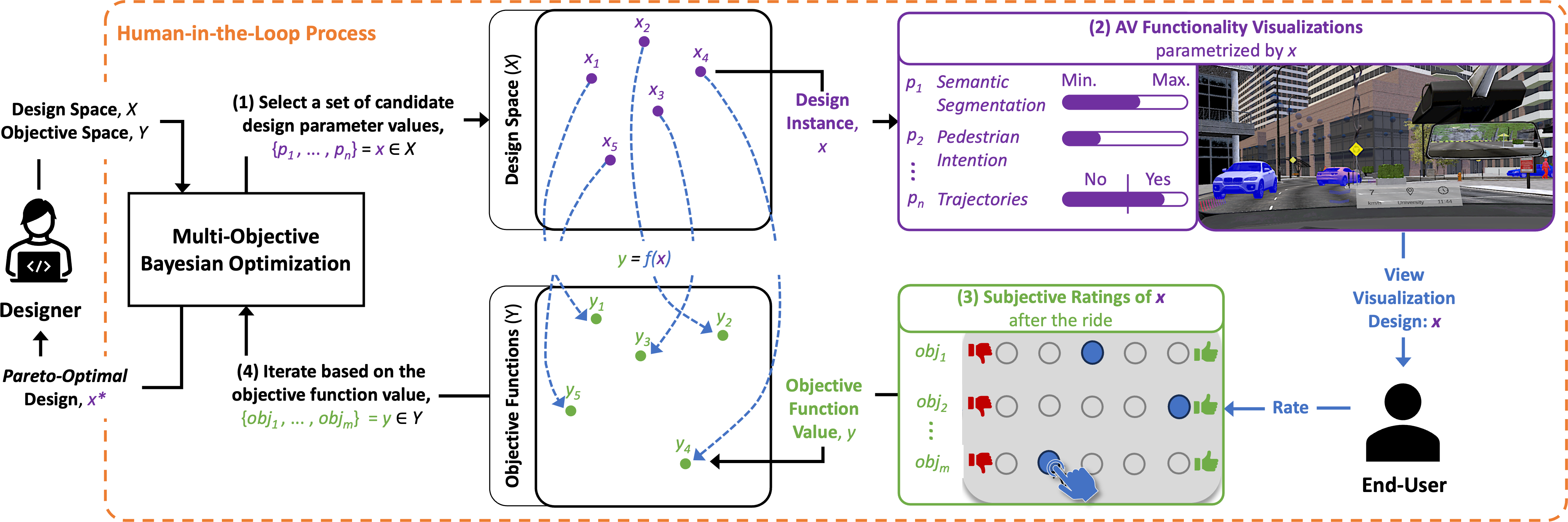}
  \caption{\opticar---human-in-the-loop multi-objective Bayesian optimization of automated vehicle (AV) functionality visualization design to increase end-users' subjective ratings of design objectives, for example, trust, perceived safety, acceptance, and aesthetics, while reducing the cognitive load ($\textit{obj}_1$ to $\textit{obj}_m$). \textbf{(1)} \opticar selects a set of parameter values (e.g., the color of semantic segmentation $p_1$ and whether to visualize vehicle trajectories $p_n$) from the design space $X$. \textbf{(2)} The end-user views the set of parameters $x$ in a simulated AV ride and \textbf{(3)} returns subjective ratings. \textbf{(4)} In the next iteration, these are used as values $y$ of the objective functions $f: X \to Y$ for which the design is optimized. Our approach finds a \textit{Pareto-optimal} \cite{marler_survey_2004} visualization design $x^*$ \textbf{per} end-user.}
  \label{fig:teaser}
  \Description{The figure shows the OptiCarVis flow chart. The Bayesian Optimization process is applied to the results, repeatedly evaluated by study participants via a questionnaire.}
\end{teaserfigure}

\maketitle

\section{Introduction}
Driving automation is anticipated to alter mobility and traffic systems~\cite{fagnant2015preparing} fundamentally. 
According to the Society of Automotive Engineers (SAE) taxonomy J3016 \cite{SAElevel}, Automated Vehicles (AVs) range from Level 4 (conditional automation) to Level 5 (full automation). As users can engage in non-driving related activities while driving tasks are automated~\cite{jansen2022design}, design priorities have extended from mainly safety concerns to user experience. 
According to ISO 9241-210~\cite{ISO9241_210_2019}, user experience incorporates all the users' emotions, beliefs, preferences, and perceptions before, during, and after use. In the AV context, user acceptance---the extent to which users are willing to use a new technology based on perceived ease of use and usefulness \cite{davis1989perceived}---contributes to high user experience~\cite{colley2021effects, colley2022scene, schneider2021explain}.

However, user acceptance of AV technology is not guaranteed. Studies have shown that many potential users are concerned about AV reliability~\cite{schoettle2014survey, kyriakidis2015public}, which refers to AVs' ability to perform driving tasks consistently safe across different situations. Both undertrust (e.g., leading to not using AVs) and overtrust (e.g., inadequately supervising the operation) present challenges for AV usage if the user's trust is inappropriate to the actual AV reliability~\cite{lee2004trust}. Therefore, prior works have shown that visualizations of AV functionality are a way to enhance user experience~\cite{colley2022scene,flohr2023prototyping,schneider2021explain}. 
Suggested visualization solutions to overcome undertrust highlighted other road users in foggy scenarios~\cite{winter2019assessing}. In the Connected Automated Driving (CAD) context, proposals include visualizing external sensor coverage (henceforth called ``CAD-covered area'') and detected road users occluded, for example, by a building (henceforth called ``occluded cars'')~\cite{mueller2022ar4cad}. Regarding overtrust, visualizing the internal AV functionalities (e.g., Situation Detection, Situation Prediction, or Trajectory Planning) and their inherent uncertainty were evaluated~\cite{colley2020effect, colley2021effects, colley2022scene, flohr2023prototyping}. 

However, the diversity of passengers complicates the design of AV functionality visualizations. Passengers' subjective perceptions of safety, trust, and aesthetics differ~\cite{mara2022acceptance}. In addition, passengers' understanding of AVs' internal functionality depends on individual knowledge and attitudes towards technology~\cite{zhang2023impact}. 
Thus, to support user understanding, designers must balance various design objectives within a complex design space (i.e., the set of possible design parameter values), for example, determining the size, transparency, and necessity of visualization elements. Traditional design methods relied on the user-centered design process (ISO 9241~\cite{ISO9241_210_2019}), standards (e.g., ISO 15005~\cite{ISO15005_2017}), and guidelines (e.g., the JAMA Guidelines for In-vehicle Display System \cite{JAMAGuideline2008}) but also experience, trial and error, and intuition especially when designing for novel AV experiences, resulting in resource-intensive user evaluations.

Prior research has explored personalization for in-vehicle displays, allowing passengers to manually adjust elements such as icon size, location, and color \cite{normark2015design}. However, translating personal preferences into effective design parameters can be difficult without design expertise. To address this, prior works introduced computational methods that rely on weighted user ratings to select from predefined visualization options \cite{yunuo2022usability, zhong2023evaluation}. Although these approaches measure usability, they omit perceived safety and trust, essential for fostering user acceptance \cite{adnan2018trust}, and restrict the design space to a small set of discrete configurations, potentially overlooking designs that better meet passengers' needs and preferences.
To improve user experience, involving humans in the design process is important to align with their needs and preferences. A Human-in-the-Loop (HITL) process \cite{chan2022bo} can iteratively present design variants to users, using their feedback to optimize design parameters. However, this optimization is challenging due to multiple objectives, such as increasing perceived safety and trust. In non-automotive User Interface (UI) domains, (Multi-Objective) Bayesian Optimization (BO and MOBO) approaches have emerged to address these design optimization problems~\cite{chan2022bo, liao2023interaction, chandramouli2023mobopersonalize, koyama2022boassistant, kadner2021adaptifont}. BO optimizes designs by predicting which parameter changes best meet the objectives. In MOBO--an extension of BO that can handle multiple objectives--manages trade-offs among conflicting objectives by identifying points on the \textit{Pareto front}, where any improvement in one objective (e.g., perceived safety through more visualizations) would lead to another objective's deterioration (e.g., increasing cognitive load). Such designs are termed \textbf{Pareto-optimal} \cite{marler_survey_2004}.

HITL MOBO can explore large design spaces in just a few iterations. However, their effectiveness in optimizing visualizations of AV functionalities is uncertain, as this is determined by subjective passenger (hereafter referred to as ''end-user'') ratings such as perceived safety and trust, which can pose a challenge for optimization (see~\cite{chandramouli2023mobopersonalize}) compared to objective measures like input error rate or accuracy (see~\cite{kadner2021adaptifont}). 
Besides, shortcomings are, for example, dealing with inconsistent human judgments~\cite{ou2022infite} and disregarding users' prior knowledge and preferences, which may reduce agency and expressiveness~\cite{chan2022bo,liao2023interaction}. 
While previous works addressed this by involving designers in HITL MOBO~\cite{chan2022bo,liao2023interaction,koyama2022boassistant}, the potential effects of including end-users without technical or design backgrounds remain largely unexplored.


To overcome these limitations, we present \opticar---the computational optimization of AV functionality visualization design using HITL MOBO. The design objectives are to increase end-users' perceived safety and trust in AV functionalities, their understanding of AVs' internal operations, and their perceived usefulness, satisfaction, and aesthetics of the visualizations while reducing their cognitive load.
We involve end-users without technical or design backgrounds in HITL optimization to leverage their valuable knowledge, experiences, and preferences. 

We demonstrate \opticar by designing visualizations for an AV’s functional levels--\textit{Situation Detection} (semantic segmentation), \textit{Situation Prediction} (pedestrian intention icons and road user trajectories), and \textit{Trajectory Planning} (AV trajectory)--along with CAD-covered areas, occluded cars, and general information (AV speed, destination, current time) to convey relevant information during automated driving stages~\cite{colley2022scene,mueller2022ar4cad,flohr2023prototyping}.
MOBO estimates the visualizations' design parameter values within the design space in each HITL iteration (see \autoref{fig:teaser}). End-users experience the selected design and provide subjective ratings \textbf{after} the ride. MOBO then uses these ratings to assess objective attainment and optimizes parameters for the next iteration, repeating until objectives are achieved (e.g., maximum trust ratings).

Our work examines six design and optimization conditions: 
We use \textbf{(1)}~No Visualization as a baseline. Additionally, to evaluate \opticar against traditional design approaches, we include \textbf{(2)}~a design created by automotive UI experts (N=8, using mean parameter values) and \textbf{(3)}~custom designs by end-users.
For our optimization conditions, we draw from previous works showing that MOBO initialized with prior data (i.e., knowledge about which design space area already achieves desired objectives; henceforth called \textit{Warm-Start}) requires fewer iterations \cite{poloczek2016warm,liao2024practical}.
Accordingly, besides \textbf{(4)}~\textit{Cold-Start} HITL MOBO with random initial design parameter values, \textbf{(5)}~we employ a \textit{Warm-Start} HITL MOBO initialized with mean parameter values from designs created by automotive UI experts (N=8).
Lastly, \textbf{(6)}~end-users create a custom design for their \textit{Warm-Start} HITL MOBO.

To quantify optimization effectiveness, end-users evaluated their final designs based on cognitive load, predictability, trust, perceived safety, usefulness, satisfaction, and aesthetics. We compared HITL MOBO results to a baseline created by automotive UI experts (N=8) and employed webcam-based eye-tracking to monitor user reactions and attention throughout the study.
In a between-subject online study with 117 participants, \opticar generated personalized visualization parameters for AVs’ \textit{Situation Detection}, \textit{Situation Prediction}, and \textit{Trajectory Planning}. These personalized designs significantly enhanced perceived safety compared to custom designs, improved predictability over expert designs, and increased trust, usefulness, and satisfaction compared to both.
Participants reported satisfaction with the design process and a sense of involvement. Many deemed the design optimal but suggested incorporating additional visualization elements and expanding driving scenarios.

By leveraging \opticar, automotive UI designers and end-users can navigate complex design spaces, potentially resulting in more passenger-centric UIs that could significantly increase their perceived safety, trust, and acceptance of AVs.
Moreover, integrating end-users' feedback into the design could inspire the development of more cooperative and effective end-user optimization methods that are implicitly integrated into future automotive UIs.

\textbf{\textit{Contribution Statement:}}
\textbf{(1)}~\opticar---the computational optimization of AV functionality visualization design using HITL MOBO to improve end-user trust, perceived safety, acceptance, and understanding of AVs and decrease cognitive load.
\textbf{(2)}~Empirical insights from a between-subject online study (N=117) evaluating \opticar against an averaged design by experts (N=8), end-user custom designs, and a No Visualization baseline. The study also examined interactions with \opticar under three optimization conditions: \textit{Cold-Start} HITL MOBO with random initial parameters, \textit{Warm-Start} HITL MOBO initialized with expert (N=8) mean parameters, and \textit{Warm-Start} HITL MOBO initialized with end-user custom parameters.
\textbf{(3)}~Open-source implementation of a Unity-based simulation enabling HITL MOBO\footnote{\url{https://github.com/Pascal-Jansen/Bayesian-Optimization-for-Unity}} of AV functionality visualizations.

\section{Background and Related Work}\label{related-work}
Our work builds on previous approaches in (1)~visualizing AV functionalities, (2)~personalization and computational methods for designing in-vehicle UIs, as well as (3)~employing HITL MOBO.

\subsection{In-Vehicle Visualizations of Automated Vehicle Functionalities}\label{rw-visualizations}



Various display technologies (e.g., HUDs, LED strips, and AR WSDs) have been evaluated for visualizing AV information. For instance, highlighting of road users in an AR WSD can reduce cognitive load \cite{colley2020effect}, while a “miniature world” view can enhance trust \cite{hauslschmid2017supportingtrust}, though users' opinion differ on its necessity. AR HUDs also affect situation awareness based on scene complexity and user driving styles, highlighting the need for personalization \cite{currano2021little}. Explaining future trajectories via an AR WSD or LED strip improves user experience, but adding a post-explanation on a smartphone confers no additional benefits \cite{schneider2021explain}. Additionally, an abstract HUD view (e.g., a symbolic icon with text) can sufficiently convey critical information (e.g., crossing children) \cite{colley2021should}.

An important factor contributing to understanding AV functionality and its safe use can be the visualization of uncertainties in detecting or predicting the driving environment and ego trajectory planning.
Anthropomorphic or abstract uncertainty indicators have been shown to boost situational awareness and trust, though they may also reduce trust if the automation appears less reliable \cite{beller2013improving, helldin2013presenting}. Due to potential increases in distraction or cognitive load, AR-based approaches have been explored instead. For instance, AR visualizations of longitudinal and lateral control uncertainties effectively convey urgency through color hues \cite{kunze2018augmented}.

Abstract visualizations (e.g., icons \cite{beller2013improving}, ambient light \cite{schneider2021explain}, miniature worlds \cite{hauslschmid2017supportingtrust}) often obscure the source of system uncertainty. In contrast, semantic segmentation of an AV’s detection process can improve situation awareness without affecting trust or cognitive load \cite{colley2021effects}. Visualizing trajectory planning can boost trust but may raise cognitive load \cite{colley2022scene}, and on-road studies confirm that showing AV functionality improves predictability, perceived usefulness, and hedonic user experiences \cite{flohr2023prototyping}. Finally, combining established designs (e.g., planned trajectories, connectivity symbols) with infrastructure support, showing occluded vehicles and merging gaps, yields the highest trust, reliability, and understanding \cite{mueller2022ar4cad}.

These previous studies highlight the individual differences among end-users in perceiving AV functionality visualizations and that these could influence their perceptions of safety, trust, acceptance, predictability, and cognitive load~\cite{currano2021little}.
As these perceptions are critical for AVs' public adoption, there is a need to align end-user experiences with AV functionality visualization design.



\subsection{Personalization and Computational Methods of In-Vehicle UI Design}\label{rw-optimization-auto-ui}
Research on automotive UIs over the past ten years~\cite{ayoub2019tenyears} and related studies~\cite{micklitz2023design} highlight the benefits of personalized in-vehicle UIs that enable tailoring UIs to end-users' perceived safety, trust, and acceptance. 
For instance, \citet{normark2015design} allowed participants to manually personalize icons' size, location, and color on the dashboard, center stack, and HUD. Normark found that participants perceived their custom designs as safer and more usable than a standard design.
However, manual personalization by end-users may be impractical. It requires dedicated design settings and is prone to human error if they (unintentionally) create inadequate designs, such as overlapping components or low-contrast colors, endangering driving safety.

Computational methods are another approach for personalization of in-vehicle UIs. These can help designers more effectively choose a design based on subjective designer and/or end-user ratings (e.g., perceived safety, trust, or acceptance).
For instance, \citet{zhong2023evaluation} derived weights from designers' usability ratings for three HUD design schemes (classic, minimalism, sport). Their computational method used these weights to select the design that best balanced the ratings. However, this method did not incorporate end-users ratings, casting doubt on whether the chosen design fully met their needs and preferences. In contrast, \citet{yunuo2022usability} allowed end-users to rate HUD design elements such as warn icon style and transparency, each at three levels. These ratings were weighted to select the most end-user-preferred HUD design from 18 predefined options. Although this method identified a design with the highest usability rating among the samples, it potentially overlooks better combinations not included in the 18 samples. Also, it does not refine the design based on iterative end-user feedback. Furthermore, these approaches \citet{zhong2023evaluation,yunuo2022usability}, focused on usability, may not adequately explore the high-dimensional design space of continuous parameter values or address multiple objectives that enhance user experience in AV functionality visualizations.

In contrast, computational optimization methods allow for iterative refinement of designs, aligning more closely with end-users' needs and preferences. Furthermore, they offer a systematic approach to personalization, potentially addressing individual preferences more effectively than manual methods (e.g., see~\cite{chan2022bo}). Despite its potential, research on optimizing in-vehicle UIs through these methods is sparse. Therefore, we use HITL optimization, which integrates designers' expertise and end-users' preferences into an iterative design optimization process. We also consider multiple design objectives, including perceived safety, trust, user acceptance, predictability, and cognitive load, exploring a wider design space with continuous parameter values.

\subsection{Human-in-the-Loop Multi-Objective Bayesian Optimization}\label{rw-hitl}\label{rw-mobo}\label{rw-comparing-designer-mobo}
HITL optimization integrates humans in its iterative parameter optimization cycles when design objectives require humans' subjective ratings (e.g.,~\cite{chiu2020human,koyama2020sequential,takagi2001interactive,zhong2021spacewalker}) or through performance measurements (e.g.,~\cite{dudley2019crowdsourcing,kadner2021adaptifont,khajah2016games}).
In Human-Computer Interaction (HCI), BO has been employed in HITL optimization to tackle various design problems~\cite{dudley2019crowdsourcing,kadner2021adaptifont,koyama2020sequential,chong2021interactive}.
BO is a machine learning method for optimizing unknown and/or difficult-to-evaluate functions~\cite{chan2022bo}, such as black-box user models.
While there are other black-box optimization methods (e.g., evolutionary and genetic algorithms, see \cite{alarie2021two}), BO stands out due to its consistent performance \cite{borji2013bayesian} and customizability \cite{liao2023human}.
BO iteratively evaluates and updates parameters to achieve the best results for a given objective. It balances \textit{exploration}, which involves probing underexplored regions of the design space to discover potentially better designs, and \textit{exploitation}, focusing on areas already identified as promising based on prior knowledge. This balance enables BO to find optimal designs with relatively few iterations, making it one of the most efficient optimization approaches \cite{brochu2010tutorial}.
Therefore, BO is well suited to the problem of AV functionality visualization design, where the relationship between design parameters and user experience is hard to model.

Previous works often use a \textit{cold-start} BO approach (e.g., \cite{chan2022bo}), where the optimization process relies on initial random sampling to gather data before refining designs. This method, while effective, can be slow due to the lack of prior knowledge \cite{liao2024practical,poloczek2016warm}. In contrast, \citet{liao2024practical} have explored a \textit{warm-start} approach, which uses pre-existing data to bypass the sampling phase, leading to faster convergence on an optimal design.

AV functionality visualization design must consider multiple objectives like safety, trust, acceptance, predictability, and cognitive load according to the Automation Acceptance Model~\cite{ghazizadeh2012extending}.
MOBO might be the answer to address these, as it can maximize or minimize multiple objectives simultaneously. The result is not a single optimal design but a range of solutions known as the Pareto front. This front consists of all designs in a multi-objective optimization problem that are not outperformed by any other design. Each point on this front is \textit{Pareto-optimal}, meaning no other design is better in all objectives simultaneously \cite{marler_survey_2004}. It illustrates the best trade-offs between conflicting objectives, where improving one (e.g., usability) may result in a decrease in another (e.g., perceived safety). 


MOBO has been used to design touchscreen keyboards balancing speed, familiarity, and spell-checking~\cite{dunlop2012multidimensional}, multi-finger mid-air text entry~\cite{dridhar2015midair}, haptic interfaces~\cite{hayward1994design}, and interactive personalization of image classifier explanations~\cite{chandramouli2023mobopersonalize}. These studies demonstrate MOBO’s effectiveness in HCI design tasks, especially when human participant studies are costly. Therefore, we argue that MOBO is suitable for optimizing AV functionality visualization design. However, its effectiveness in this domain remains unclear, particularly in modeling complex end-user states such as perceived safety, trust, and acceptance.

Besides, \citet{chan2022bo} and \citet{liao2023interaction} revealed that designers felt less agency and ownership over MOBO-driven designs, even if they were of higher quality. To foster collaboration between BO-supported design approaches and designers, \citet{koyama2022boassistant} introduced BO as a design assistant. This allowed designers to leverage expertise and preferences while BO offered design suggestions.

In contrast, involving end-users in AV functionality visualization design is critical, as they offer key insights into their preferences that designers cannot have. While prior research primarily explored the impact of designers in HITL MOBO design processes (and vice versa), end-user integration is left underexplored. Yet, their contributions may enhance optimization efficiency and design quality compared to designers. Therefore, this work focuses on engaging end-users in the design of visualizations to address existing HITL MOBO limitations. Besides, we extend the understanding of qualities of HITL optimization established in works like~\cite{chan2022bo,liao2023interaction} by comparing end-user-led and optimizer-driven processes.

\section{\opticar: Optimizing Automated Vehicle Functionality Visualizations}\label{method-opticar}
Visualizations on WSDs use icons~\cite{lindemann2018catch}, highlighting~\cite{colley2021effects,jansen2024visualizing}, and other elements (e.g., see~\cite{kunze2019conveying}) to communicate AV functionalities, which could be crucial for user acceptance~\cite{colley2021effects,jansen2024visualizing}. 
These visualizations aim to increase end-users' perceived safety and trust in AVs, their understanding of AV actions in various driving situations (e.g., unexpected stopping) \cite{jansen2024visualizing}, and reduce the cognitive load when processing this (potentially overwhelming) information. 

\begin{figure*}[h]
    \centering
    \includegraphics[width=\linewidth]{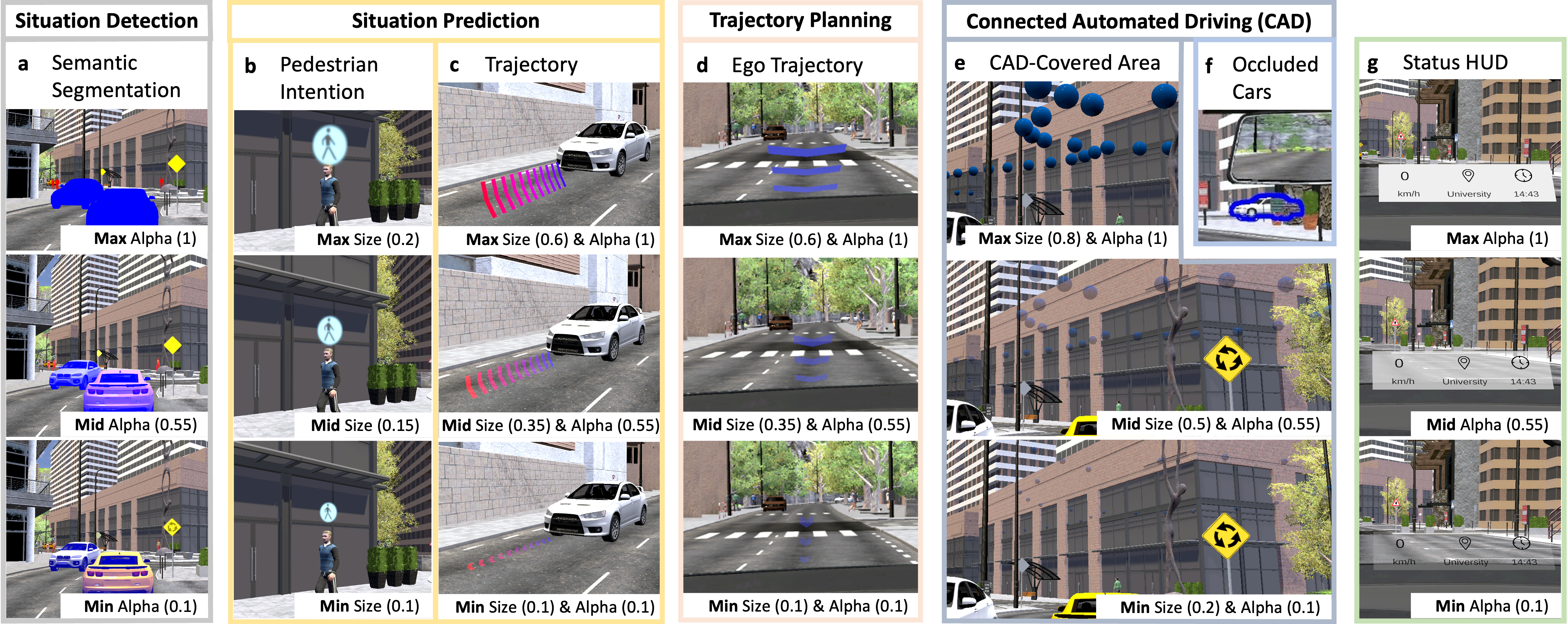}
    \caption{Overview of the employed visualizations of an SAE Level 4 \cite{SAElevel} AV's functional levels of internal operation \cite{colley2022scene}, CAD \cite{mueller2022ar4cad}, and status on an AR WSD, showing the possible variations in transparency (alpha) and size values (see brackets). \textit{Min} and \textit{Max} represent the designs at the lower and upper bounds of the continuous parameter ranges, while \textit{Mid} represents the midpoints.}
    \label{fig:in-vehicle-visualizations-overview}
    \Description{The figure gives an overview of six different visualization concepts by showcasing them in action. For each concept, three variations display different levels of transparency and size. Shown are the trajectory, pedestrian intention, semantic segmentation, occluded cars, ego trajectory, CAD-covered area, and car status in HUD.}
\end{figure*}

We aim to optimize the visualization design computationally using a HITL MOBO approach. In this work, we solely focus on AR WSD visualizations as these allow situated visualizations, which are beneficial for understanding and reducing cognitive load~\cite{colley2020effect}.
By involving end-users, the HITL MOBO process can iteratively refine designs to cater to individual preferences within a reasonable timeframe. 
Our visualizations (see \autoref{fig:in-vehicle-visualizations-overview} and \autoref{fig:driving_route}) are grounded in prior research investigating trust, cognitive load, and perceived safety in AVs (see Section \ref{rw-visualizations}). We primarily built upon the work of \citet{colley2022scene}, who visualized the functional levels of AVs' internal operations (see~\cite{dietmayer2016predicting}): \textit{Situation Detection}, \textit{Situation Prediction}, and \textit{Trajectory Planning}.

Within this framework, Situation Detection is encoded via semantic segmentation of detected objects (i.e., detected vehicles are colorized in blue, pedestrians in red, and traffic signs in yellow, see \autoref{fig:in-vehicle-visualizations-overview} a).
Situation Prediction is encoded via showing the pedestrian intentions symbolized as icons above pedestrians' heads~\cite{colley2020effect} (i.e., the color coding indicates the prediction whether pedestrians are to cross the street (dark blue), whether they will remain on the sidewalk (cyan), or whether the prediction is uncertain (yellow); see \autoref{fig:in-vehicle-visualizations-overview} b), and deduced trajectories of other vehicles are shown as a line. Here, the color changes from blue to red the further the prediction lies in the future to visualize the increasing uncertainty. The Trajectory Planning~\cite{colley2022scene} (i.e., the planned trajectory of the own vehicle) is also visualized via this line (see \autoref{fig:in-vehicle-visualizations-overview} c and d).
Furthermore, we incorporated elements from \citet{mueller2022ar4cad}, focusing on CAD visualizations. CAD supports the AV in all three framework stages and includes additional information as blue spheres above the road, indicating the AV's active link to external sensors in a given area (see  \autoref{fig:in-vehicle-visualizations-overview} e). We also include an outline representing occluded cars through buildings (see \autoref{fig:in-vehicle-visualizations-overview} f). This visualization supports the end-user in understanding that the AV knows about other vehicles even if they are hidden, for example, by buildings, which may be unintuitive~\cite{mueller2022ar4cad}.
Finally, a vehicle status HUD displays the current time, AV speed, and destination as basic information (see \autoref{fig:in-vehicle-visualizations-overview} g).

\label{method-bo}
We define the optimization of AV functionality visualization design as the task of finding the design parameter combination $x* = \{p_1, \dots, p_n\}$ such that:
\begin{equation} \label{argmax-equation}
x* = \argmax_{x \in X} f(x)
\end{equation}
where $X \subseteq \mathbb{R}^n$ is the design space defined by $n$ design parameters $p \in \mathbb{R}$. The objective function $f: X \to Y \subseteq \mathbb{R}^m$ maps each design $x$ to $m$ subjective metrics (e.g., trust and perceived safety). An objective function value $y \in Y$ is a subjective metric rating (e.g., via Likert scale) the end-user returns to the MOBO in the HITL process after viewing a visualization design $x$ (see \autoref{fig:teaser}). As the relationship between $x$ and $y$ is unclear, we define this as a black-box function $y = f(x)$~\cite{alarie2021two}. 

\subsection{Design Parameters}\label{method-design-params}
The design parameters for the visualizations were derived from the respective publications~\cite{kunze2018augmented, colley2020effect, colley2022scene, mueller2022ar4cad, currano2021little, riegler2019adaptive, 10.1145/3631408}.
As visualization elements might be unwanted, we defined the visualization visibility $v$ per element as $v \in [0, 1]$.
Specifically, we set a threshold to map $v$ as a Boolean. An element is invisible for $v < 0.5$ and visible for $v \geq 0.5$. We employed this mapping as BO is typically more efficient with continuous parameters~\cite{shahriari2015taking}. This Boolean value was determined for the semantic segmentation, (ego) trajectory, pedestrian intention, highlighting occluded cars, the CAD-covered area, and the vehicle status HUD.
Besides, element size may denote importance and determine far-distance visibility. Therefore, we added the size $s$ as a parameter of the pedestrian intention icon, (ego) trajectory, and CAD-covered area sphere. We defined $s$ within a range where the bounds indicate the smallest and the largest appropriate size. These bounds were different for each element so that they can neither be too small and thus invisible nor too large and overlap with other elements.
Due to AR WSD elements potentially overlaying the driving environment, like other vehicles, their semi-transparency might increase the visibility of the environment. We assigned an alpha level $\alpha$ to the semantic segmentation, (ego) trajectory, CAD-covered area sphere, and vehicle status HUD. The range was $\alpha \in [0.1, 1]$ as elements become nearly invisible for $\alpha < 0.1$. The ''occluded cars'' visualization does not incorporate an alpha value because it does not block relevant visual information (''occluded cars'' is a simple outline). Besides, we did not assign an $\alpha$ level to the pedestrian intention icon to avoid confusion between color coding because semi-transparent dark blue (likely to cross) looks similar to cyan (remaining on the sidewalk).

We avoided RGB coloring as parameters as the color was already chosen not to convey unintended meaning (e.g., orange being a warning signal; e.g., see~\cite{colley2022scene, mueller2022ar4cad}).
We also refrained from altering the position of the CAD-covered area spheres (e.g., via height) and the vehicle status HUD (e.g., x and y position on the windshield) as the proposed constant positions are the most useful.
During optimization, these constant positions prevent visualization elements' misalignment (e.g., overlapping) and ensure visibility for multiple passengers' viewpoints.
Similarly, we disregard the visualization elements' rotations as these are already determined by the objects' orientations in the environment, such as vehicles and pedestrians.
All design parameters ($p_1$ to $p_{16}$) are summarized in \autoref{tab:design_param}.

\begin{table*}[ht!]
\scriptsize
\caption{The 16 design parameters for the visualization design, with the ranges in Unity values. All design parameters are modeled continuously, with values mapped to Boolean if necessary (``Bool''). Example visualizations of parameter values are shown in \autoref{fig:in-vehicle-visualizations-overview}.}
\label{tab:design_param}
\resizebox{\textwidth}{!}{%
\begin{tabular}{@{}llll@{}}
\toprule
\textbf{Design Parameter }                  & \textbf{Description}                                  & \textbf{Reference}                 & \textbf{Range}                  \\ \midrule
$p_1$: Semantic Segmentation, $v$       & Whether the semantic segmentation result should be visualized.   &~\cite{colley2021effects}  & [0, 1]; Bool \\
$p_2$: Semantic Segmentation Alpha, $\alpha$ & Alpha value of the semantic segmentation.                        &~\cite{colley2021effects}  & [0.1, 1]             \\ \hdashline

$p_3$: Pedestrian Intention, $v$        & Whether the predicted pedestrian intention should be visualized. &~\cite{colley2020effect}   & [0, 1]; Bool \\
$p_4$: Pedestrian Intention Size, $s$   & Alpha value of the pedestrian intention symbol.                  &~\cite{colley2020effect}   & [0.1, 0.2]             \\ \hdashline

$p_5$: Trajectory, $v$                  & Whether the predicted trajectory of others should be visualized. &~\cite{kunze2018augmented, 10.1145/3631408} & [0, 1]; Bool \\ 
$p_6$: Trajectory Alpha, $\alpha$            & Alpha value of the trajectory.                                   &~\cite{kunze2018augmented} & [0.1, 1]             \\
$p_7$: Trajectory Size, $s$             & Size of the trajectory.                                          &~\cite{kunze2018augmented} & [0.1, 0.6]             \\ \hdashline

$p_8$: Ego Trajectory, $v$   & Whether the own planned trajectory should be visualized.                  &~\cite{colley2022scene, 10.1145/3631408}   &  [0, 1]; Bool         \\ 
$p_{9}$: Ego Trajectory Alpha, $\alpha$   & Alpha value of the own planned trajectory.                  &~\cite{colley2022scene, 10.1145/3631408}   & [0.1, 1]              \\ 
$p_{10}$: Ego Trajectory Size, $s$             & Size of the own planned trajectory.                                          &~\cite{kunze2018augmented} & [0.1, 0.6]             \\ \hdashline

$p_{11}$: CAD-Covered Area, $v$                & Whether the area covered through V2x should be visualized.       &~\cite{mueller2022ar4cad}  & [0, 1]; Bool \\
$p_{12}$: CAD-Covered Area Alpha, $\alpha$         & Alpha value of the symbols for the CAD-covered area.                     &~\cite{mueller2022ar4cad}  & [0.1, 1]    \\    
$p_{13}$: CAD-Covered Area Size, $s$         & Size of the symbols for the CAD-covered area.                     &~\cite{mueller2022ar4cad}  & [0.2, 0.8]    \\ \hdashline

$p_{14}$: Occluded Cars, $v$              & Whether occluded (e.g., by buildings) cars should be visualized. &~\cite{mueller2022ar4cad}  & [0, 1]; Bool \\ \hdashline

$p_{15}$: Vehicle Status HUD, $v$              & Whether the vehicle status in the HUD should be visualized.      &~\cite{currano2021little}                          & [0, 1]; Bool \\
$p_{16}$: Vehicle Status HUD Alpha, $\alpha$            & Alpha value of the vehicle status.       &~\cite{riegler2019adaptive}                & [0.1, 1]\\ \bottomrule
\end{tabular}%
}
\end{table*}

\subsection{Objective Functions}\label{method-objectives}
An objective function $f$ maps a visualization design $x$ to a subjective metric the optimizer seeks to maximize or minimize with the design.
According to our optimization goal, we consider five subjective metrics - \textit{safety}, \textit{trust}, \textit{predictability}, \textit{acceptance}, and \textit{aesthetics} - to be maximized. \textit{Cognitive load} was our sole subjective metric to be minimized.

Based on previous work~\cite{colley2022scene,colley2020effect}, we employed the following questionnaires to retrieve these metrics after every optimization iteration in the HITL process:
We assessed \textbf{cognitive load} via the mental workload subscale of the raw NASA-TLX~\cite{hart1988development} on a 20-point scale (``How much mental and perceptual activity was required? Was the task easy or demanding, simple or complex?''; 1=\textit{Very Low} to 20=\textit{Very High}; lower is better).
Regarding predictability and trust, we used the subscales \textit{Predictability/Understandability} (\textit{Predictability}) and \textit{Trust} of the \textit{Trust in Automation} questionnaire by \citet{korber2018theoretical}.
\textbf{Predictability} is determined via agreement on four statements (``The system state was always clear to me.'', ``I was able to understand why things happened.''; two inverse: ``The system reacts unpredictably.'', ``It's difficult to identify what the system will do next.'') using 5-point Likert scales (1=\textit{Strongly disagree} to 5=\textit{Strongly agree}).
\textbf{Trust} is measured via agreement on the same 5-point Likert scale on two statements (``I trust the system.'' and ``I can rely on the system.''; both times, higher is better).
Participants rated their perceived \textbf{safety} using four 7-point semantic differentials from -3 (anxious/agitated/unsafe/timid) to +3 (relaxed/calm/safe/confident; higher is better)~\cite{faas2020longitudinal}.
Finally, we added three single items.
Two were defined with the van der Laan acceptance scale~\cite{van1997simple} in mind (``I find the visualizations of the automated vehicle \textbf{useful}'', ``I find the visualizations of the automated vehicle \textbf{satisfying}''). These were combined into a single ``acceptance'' objective.
We also adapted the question regarding \textbf{aesthetics} from \citet{colley2023uam} (``I found the visualizations visually appealing''; on a 7-point Likert scale).

Normalization is required before submitting these to the optimizer because the subjective metrics values have ranges based on 20-, 5-, or 7-point Likert scales.
We transformed these six metrics into the $[-1, 1]$ range.
After this, the \textit{cognitive load} objective is a function to be maximized (a higher value means less load).


\subsection{Hyperparameter Setup for Bayesian Optimization}\label{method-hyperparam-setup}
For our MOBO implementation, we used the PyTorch-based library \texttt{BoTorch}~\cite{balandat2020botorch} in version 0.9.2.
As we have a multi-objective setup, we employed the multi-output Gaussian Process and applied \texttt{qEHVI} as the acquisition function.
This function represents the expected hypervolume increase, where we set q = 1 (in line with~\cite{chan2022bo}) to ensure that after each iteration, a batch of size one is selected to be given to the end-user for evaluation.
Other hyperparameter settings were 5 sampling iterations followed by 10 optimization iterations. During the optimization of the acquisition function, 2024 restart candidates for the acquisition function optimization, and 512 Monte Carlo samples were used to approximate the acquisition function.
These settings are based on \citet{chan2022bo}. 

\label{method-stop-criterion}
In internal tests, we found that the convergence to an optimal rating of the objectives was reached rather quickly.
Therefore, we added a stopping criterion checked after every measurement: Was the perfect rating for \textbf{every} subjective metric (i.e., the \textbf{highest} rating for trust, predictability, safety, aesthetics, usefulness, satisfaction, and the \textbf{lowest} rating for cognitive load; see Section \ref{method-objectives}) given for the \textbf{last} round? Participants could otherwise not opt out of the optimization steps.

\section{Experiment}\label{experimental-method}
We aim to empirically validate the effectiveness of the HITL MOBO approach for designing AV functionality visualizations, comparing it with traditional manual designs. Building on \citet{chan2022bo}, we investigate the mutual influence between the HITL optimization and its end-users during simulated automated driving. Additionally, we expect varied outcomes when initializing the MOBO (i.e., Warm-Start) with data from automotive UI experts or end-users. Guided by these goals, we conducted a user study with the following research questions (RQs):
\begin{itemize}
    \itema How does the HITL MOBO of AV functionality visualizations impact end-users' rating of safety, trust, predictability, acceptance, aesthetics, and cognitive load?
    \itemb Which condition produces the design leading to the highest rating of safety, trust, predictability, acceptance, aesthetics, and the lowest cognitive load: Cold-Start with random initial parameters, Warm-Start initialized by expert designs, or Warm-Start initialized by end-user designs?
    \itemc How does the participation in a HITL optimization process affect end-users, and what are their areas of interest during design?
\end{itemize}

The experimental procedure followed the guidelines of our university's ethics committee and adhered to regulations regarding handling sensitive and private data, anonymization, compensation, and risk aversion. Compliant with our university's local regulations, no additional formal ethics approval was required.


\subsection{Apparatus}\label{study-apparatus}
The apparatus comprises three main components: (1) a Unity application for the driving environment, (2) a custom parameter design tool, and (3) a Bayesian optimizer. We also used a server to persistently store the local questionnaire responses persistently, the design parameter logs from optimization rounds, participants' ratings, and optimization durations.

\subsubsection{Automated Vehicle and Driving Environment}\label{study-apparatus-unity}
\begin{figure*}[ht!]
\centering
    \includegraphics[width=\linewidth]{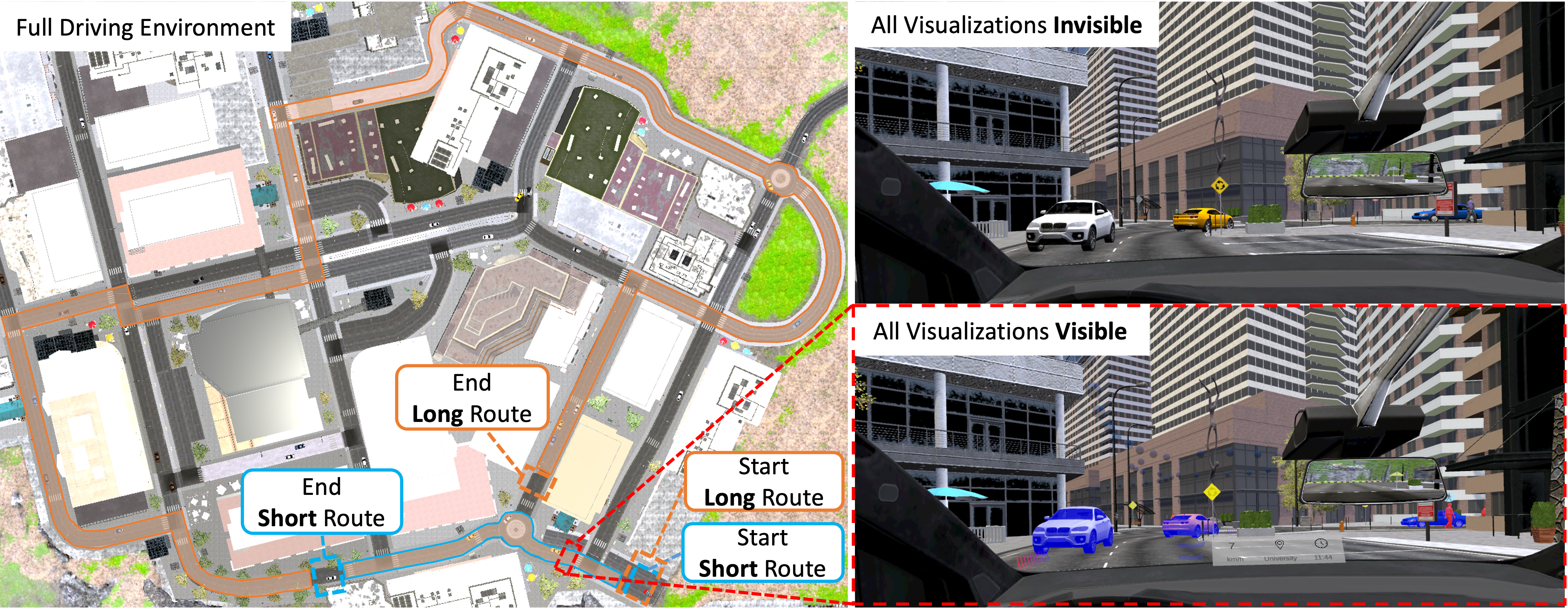}
   \caption{AV study driving route used in the HITL MOBO iterations (blue) and long route used in the final assessments (orange). Besides, examples of the driver's perspective with all visualizations visible using \textit{mid} transparency and size values (red).}
   \label{fig:driving_route}
    \Description{The figure shows the driving route of the study, with the start, end, and critical driving situations highlighted at the road segments. Besides, two screenshots of the driving scene are from the driver's perspective. One with all visualizations visible and one with invisible visualizations.}
\end{figure*}

We designed a standalone application for Windows and macOS using \href{https://unity.com/}{Unity} 2022.3.7. This application simulates in-vehicle visualizations in a driving environment using a 3D model of the Tesla Model X, modified to feature a virtual AR WSD and a vehicle status HUD. As participants should not overtake control but expect errors in automation, we consider a SAE Level 4 AV \cite{SAElevel}. The driving environment, Unity Windridge City, aligns with previous research~\cite{colley2020effect,colley2021effects,colley2022scene}. We used the \href{https://assetstore.unity.com/packages/templates/systems/urban-traffic-system-89133}{Urban Traffic System} asset to simulate pedestrian and vehicle behaviors. Each MOBO iteration employs a fixed 33-second route. This brief duration is chosen to reduce user fatigue during the HITL process. We also developed a longer 3-minute route to provide users with a broader range of traffic situations. Both routes (see \autoref{fig:driving_route}) incorporate frequent pedestrian and vehicle interactions at roundabouts and zebra crossings, creating diverse visualization scenarios.

\begin{figure*}[ht!]
\centering
    \includegraphics[width=\linewidth]{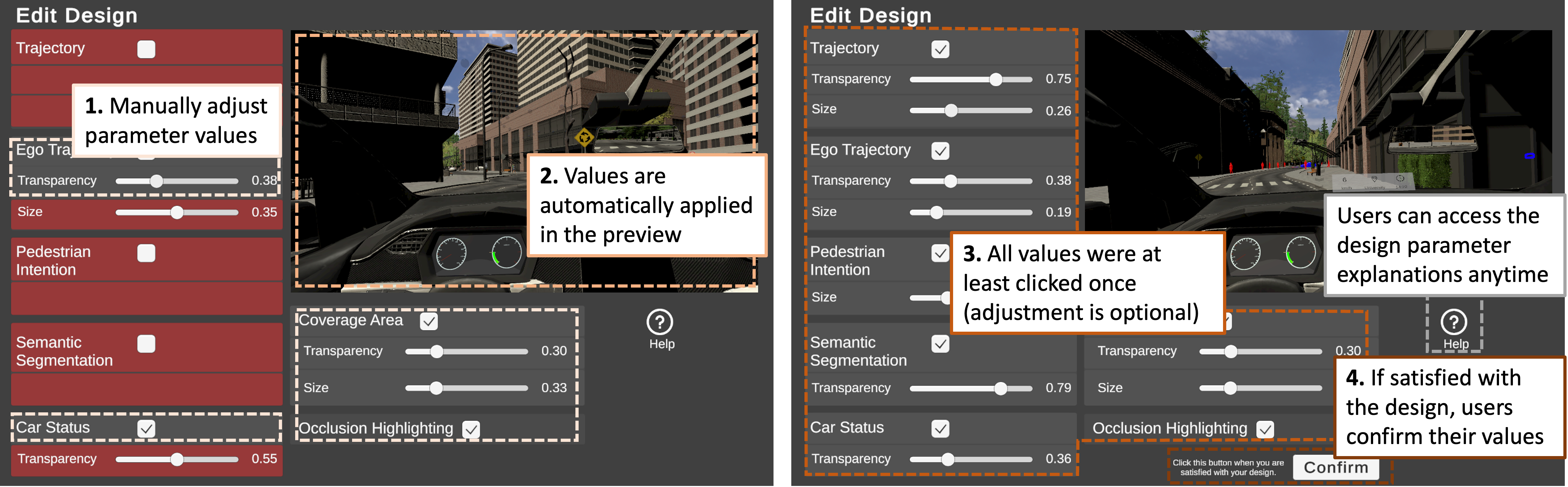}
   \caption{The custom parameter design tool allows for adjusting 16 parameters (see \autoref{tab:design_param}). \textbf{(1)} Users modify values using checkboxes and sliders, with untouched settings highlighted in red. \textbf{(2)} The adjusted values are displayed in a preview that loops the AV driving environment. \textbf{(3)} After interacting with all settings once (their adjustment is optional), \textbf{(4)} the ''confirm'' button activates. The parameter explanation view (see \autoref{fig:intro}) is accessible via the ''help'' button.}
   \label{fig:param-design-tool}
    \Description{The figure shows the custom parameter design tool in two screenshots. In the left picture, the first step of the custom design process is shown, including the red highlighting if a parameter setting has not yet been used. The preview window shows the driving scene from the driver's perspective. The right picture shows the final step when all parameter settings were used once, and the confirm button on the lower right side appeared.}
\end{figure*}

\subsubsection{Custom Parameter Design Tool}\label{study-apparatus-design-tool}
In Unity, we developed a tool (see \autoref{fig:param-design-tool}) that enables adjustment of the 16 parameters (see \autoref{tab:design_param}) for custom visualization designs. Users can toggle the element visibility $v$ using a checkbox and adjust the element transparency $\alpha$ and size $s$ using sliders within the predefined parameter value ranges. A side-by-side preview panel continuously displays the AV driving environment with the current settings. As users modify parameters, this environment loops. Once users finalize their settings, they confirm via a button, saving the parameter value configuration as initial data for the Bayesian optimizer.

\subsubsection{The Bayesian Optimizer}\label{study-apparatus-bo}
During the HITL optimization, the Bayesian optimizer interacts with the Unity application. It iteratively receives user ratings for the current visualization design (the optimization objectives, see Section \ref{method-objectives}) and returns the next potentially optimal parameters in CSV format. To guarantee prompt computation, the optimizer runs locally on participants' computers. 
The configuration of this optimizer for in-vehicle visualization design is detailed in Section~\ref{method-hyperparam-setup}.

\subsection{Design and Optimization Conditions}\label{study-optimization-strategies}
To answer RQ1 - RQ3 (see Section \ref{experimental-method}), we employ the following conditions, building upon prior research~\cite{chan2022bo,liao2023interaction}:

\begin{itemize}

    \itemas \textbf{No Visualization (No Vis.): } In this condition, no visualization of AV functionalities is shown.

    \itembs \textbf{Custom design by experts: }
End-users evaluate a \textit{standard} design created by automotive UI experts (N=8, see Section \ref{expert-study}) using our parameter design tool (see Section~\ref{study-apparatus-design-tool} and \autoref{tab:expert}). This visualization uses the mean parameter values from all expert designs. Unlike C5, this condition may result in suboptimal designs, as a "one-size-fits-all" \textit{standard} expert design alone may not optimally adhere to the individual subjective ratings of end-users.

    \itemcs \textbf{Custom design by end-users: }
Similar to C2, but instead of evaluating a \textit{standard} design, end-users manually personalize visualizations using our parameter design tool and evaluate them after the AV ride. Unlike C6, this condition may result in suboptimal designs, as end-users may not fully understand their preferences and ineffectively translate them in a direct parameter design process.

    \itemds \textbf{Cold-Start HITL MOBO: }
We use a Cold-Start HITL MOBO initialized with random parameters generated by the optimizer. The MOBO starts in the \textit{sampling} phase. End-users then interactively rate potential designs, fed back into the optimizer in a HITL process.

    \itemes \textbf{Expert-Informed Warm-Start HITL MOBO: }
As in C2, we enable automotive UI experts (N=8, see Section \ref{expert-study}) to explore designs using our parameter design tool. End-users then rate this \textit{standard} expert design, which creates design objective values for the given parameters. In the Warm-Start approach, these values initialize the HITL MOBO. Thus, the \textit{sampling} phase is replaced, and it directly starts with the \textit{optimization} phase, in which the end-user is iteratively involved. This condition leverages experts' domain knowledge to focus the design space on areas likely aligned with automotive UI best practices. It potentially optimizes designs by further personalizing a given design foundation to fit end-users' subjective ratings.

    \itemfs \textbf{User-Informed Warm-Start HITL MOBO: }
End-users first explore personalized visualization designs using our parameter design tool. After that, they rate their design to generate the objective values to initialize the HITL MOBO \textit{optimization} phase. This Warm-Start HITL MOBO process then fine-tunes the user-informed parameter values, combining end-users' preferences with optimization to uncover optimized designs when they cannot fully express their needs in a custom design. Like in C5, this potentially accelerates MOBO's discovery of optimal designs as end-user preferences narrow the design space~\cite{chan2022bo,liao2023interaction}. Besides, such Warm-Start HITL MOBO augmentation could combat users' feelings of low agency in HITL approaches~\cite{chan2022bo}.

\end{itemize}

Only C4-C6 represent personalization methods that include HITL MOBO. C1-C3 represent the current state of the art by not showing any information or letting experts or end-users define the visualization themselves.
We exclusively employ MOBO as our black-box optimization method due to its consistently robust performance across applications \cite{borji2013bayesian}, unparalleled customizability tailored to the unique requirements of HCI design \cite{liao2023human}, and superior efficiency in converging on optimal designs \cite{brochu2010tutorial}. Therefore, we argue that including competitor optimization methods would unlikely add value to our experiment.
We define the following hypotheses:
\begin{itemize}
    \itemha HITL MOBO for AV functionality visualizations (C4-C6) will increase end-users ratings of safety, trust, predictability, acceptance, and aesthetics and reduce cognitive load compared to non-MOBO conditions (C1-C3).
    \itemhb Among the HITL MOBO conditions, the C6-User-Informed Warm-Start will result in higher ratings for safety, trust, predictability, acceptance, aesthetics, and lower cognitive load, outperforming both the C4-Cold-Start and the C5-Expert-Informed Warm-Start.
\end{itemize}



\subsection{Expert Study to Inform the Standard Visualization Design}\label{expert-study}
For conditions C2 and C5, addressing RQ2, we aimed to create a \textit{standard} visualization design using expert insights on typical automotive UI design practices. We recruited N=8 automotive UI experts (2 female, 6 male, 0 non-binary) who specialized in in-vehicle UI usability and trust in automation. These experts, with backgrounds in psychology (1), computer science/HCI (6), and engineering (1), represented four institutions from Europe, the USA, and Canada. They hold positions as research associates and Ph.D. students or are currently or were engineers at two large European OEMs. Participants were, on average, \m{27.88} (\sd{2.36}) years old. All have published multiple papers on automotive design. Publishing in automotive design-oriented venues constitutes \textit{Experience} and \textit{Peer Identification}, which, in the sense of \citet{shanteau2003can}, constitute experts. The participants' expertise in designing \textbf{and} evaluating automotive UIs regarding subjective end-user ratings makes them a perfect fit for this study. 
The experts were tasked to design AR WSD visualizations for AVs that enhanced end-users' perceived safety, trust, predictability, and acceptance of AVs while reducing cognitive load.

\begin{figure*}[ht!]
\centering
    \includegraphics[width=\linewidth]{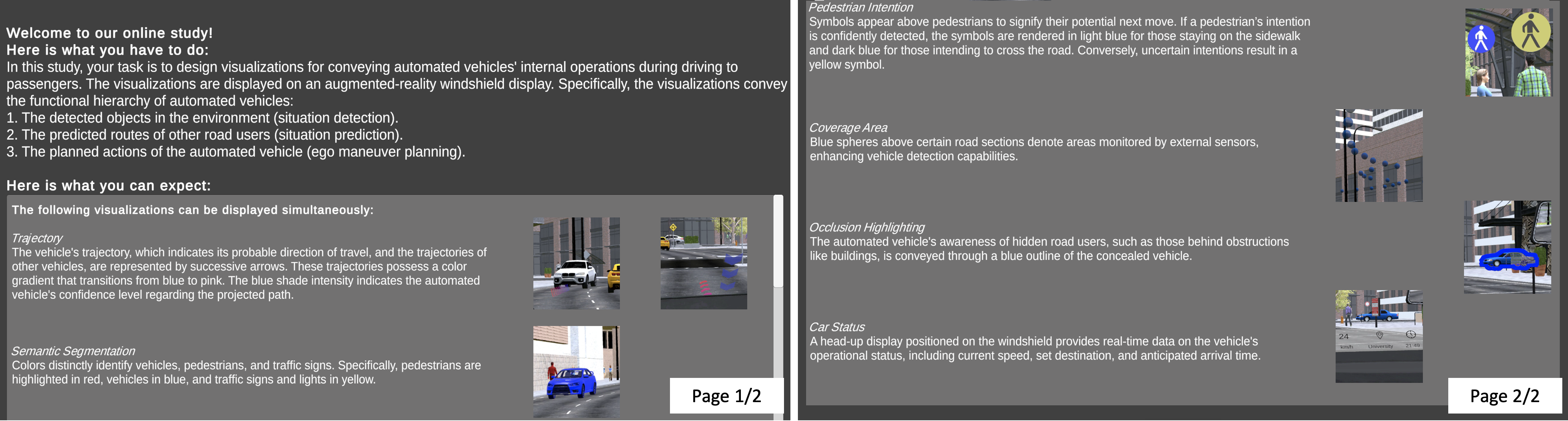}
   \caption{Excerpt of the information given to study participants at the start. Participants were also questioned about the visualizations to ensure understanding.}
   \label{fig:intro}
    \Description{The figure shows two screenshots of the initial information given to study participants. Participants were greeted with a brief introduction about the study task and short descriptions of the six visualization concepts: trajectory, pedestrian intention, semantic segmentation, occluded cars, CAD-covered area, and car status.}
\end{figure*}

Each session started with a brief instruction on the available visualizations (see \autoref{fig:intro}) and parameters (see \autoref{tab:design_param}), informed consent, and a demographic questionnaire. Using the custom design tool in Unity (see Section~\ref{study-apparatus-design-tool} and \autoref{fig:param-design-tool}), experts freely adjusted the 16 design parameters. By default, the visibility checkboxes were set to false, while the sliders were set to the mid-value. The preview panel continuously visualized the current configuration in a looped scene, enabling experts to refine their designs iteratively. Once satisfied, they confirmed their parameter configuration and answered open-ended questions about design rationales. The resulting parameter values are shown in \autoref{tab:expert}. 
For C2 and the initialization of the HITL MOBO in C5, we used a \textit{standard} visualization derived by averaging each parameter from the expert designs.
The averaging of expert opinions avoids bias towards any viewpoint and balances contrasting parameters (e.g., a trajectory alpha of one compared to 0.31).

To verify the averaged design, three authors individually reviewed and then collaboratively discussed it, supplemented by qualitative comparisons to the previous works from which we derived our visualizations \cite{kunze2018augmented, colley2020effect, colley2022scene, mueller2022ar4cad, currano2021little, riegler2019adaptive}. 
Most design parameters exhibited relatively limited variability ($SD \leq 0.18$; see \autoref{tab:expert}). Specifically, Semantic Segmentation ($p_1, p_2$), Pedestrian Intention ($p_3, p_4$), Trajectory ($p_5, p_7$), Ego Trajectory Alpha ($p_9$) and Size ($p_{10}$), CAD-Covered Area Size ($p_{13}$), and Vehicle Status HUD ($p_{15}$) had $SD \leq 0.18$, indicating a fair degree of alignment among experts. In contrast, six parameters had more pronounced variability ($SD > 0.18$), suggesting diverse views. These included Trajectory Alpha ($p_6$), Ego Trajectory ($p_8$), CAD-Covered Area ($p_{11}, p_{12}$), Occluded Cars ($p_{14}$), and Vehicle Status HUD Alpha ($p_{16}$).
Post-study interviews indicated that personal preferences influenced alpha parameter choices, making it challenging for experts to settle on a general design suited for all users. However, the comparatively lower variability across most parameters suggests the averaged design parameters are reasonable. After reviewing the final averaged design, all experts noted it fell within an acceptable range of values. This combination of quantitative averaging and qualitative insights aligns with established industry practices for consensus building in empirical A/B studies~\cite{mattos2020automotive}.

\subsection{Participants}\label{study-participants}
We computed the desired sample size for the main experiment via an a-priori power analysis using G*Power \citep{faul2009statistical}.
To achieve a power of 0.95, with an alpha of 0.05, 111 participants should allow for detection of a medium effect (\textit{Effect Size f}=0.25) in repeated measures ANOVA with the conditions C1-C6 as a between-subject factor.

Thus, we recruited 117 participants (Mean age = 39.2, SD = 12.3, range: [19, 72]; Gender: 29.9\% women, 68.4\% men, 1.71\% non-binary; Education: College, 72.65\%; High School, 20.51\%; Vocational training, 6.84\%) via \url{prolific.co}. 
C1 had 21 participants, C2 19, C3 19, C4 18, C5 22, and C6 18.

To prevent confounding effects of traffic handedness (right-hand vs. left-hand traffic) or culture, the participant pool was limited to US residents~\cite{rasouli2019autonomous} as the Unity simulation employed right-handedness.
Regarding their employment status, 81 are employees, eight are students at a college, one is at a school, 15 are self-employed, 9 are job-seeking, and three indicated \textit{other}.
All participants hold a valid driver's license for, on average, \m{18.62} (\sd{13.18}) years. 
We found no significant differences between the \optimization for license, gender, or age.
All volunteered under informed consent and agreed to the recording and anonymized publication of results.
Participants were compensated with £7.

\subsection{Procedure}\label{study-procedure}
We conducted the study online to engage diverse end-users with non-technical backgrounds, a typical challenge in lab settings. Moreover, safely simulating AVs in a computer-screen-based Unity application is a widely adopted method for evaluating novel in-vehicle UIs (e.g., see~\cite{jansen2022design, colley2020effect, colley2022scene}) as such an AV technology is not yet available.
Using a between-subjects, the study employed the distinct conditions \textbf{C1-C6}. 
The baseline condition (C1-No Vis.), which displayed no visualizations during the AV ride, was added to validate the effectiveness of designs from C2-C6.
\begin{figure*}[ht!]
\centering
    \includegraphics[width=\linewidth]{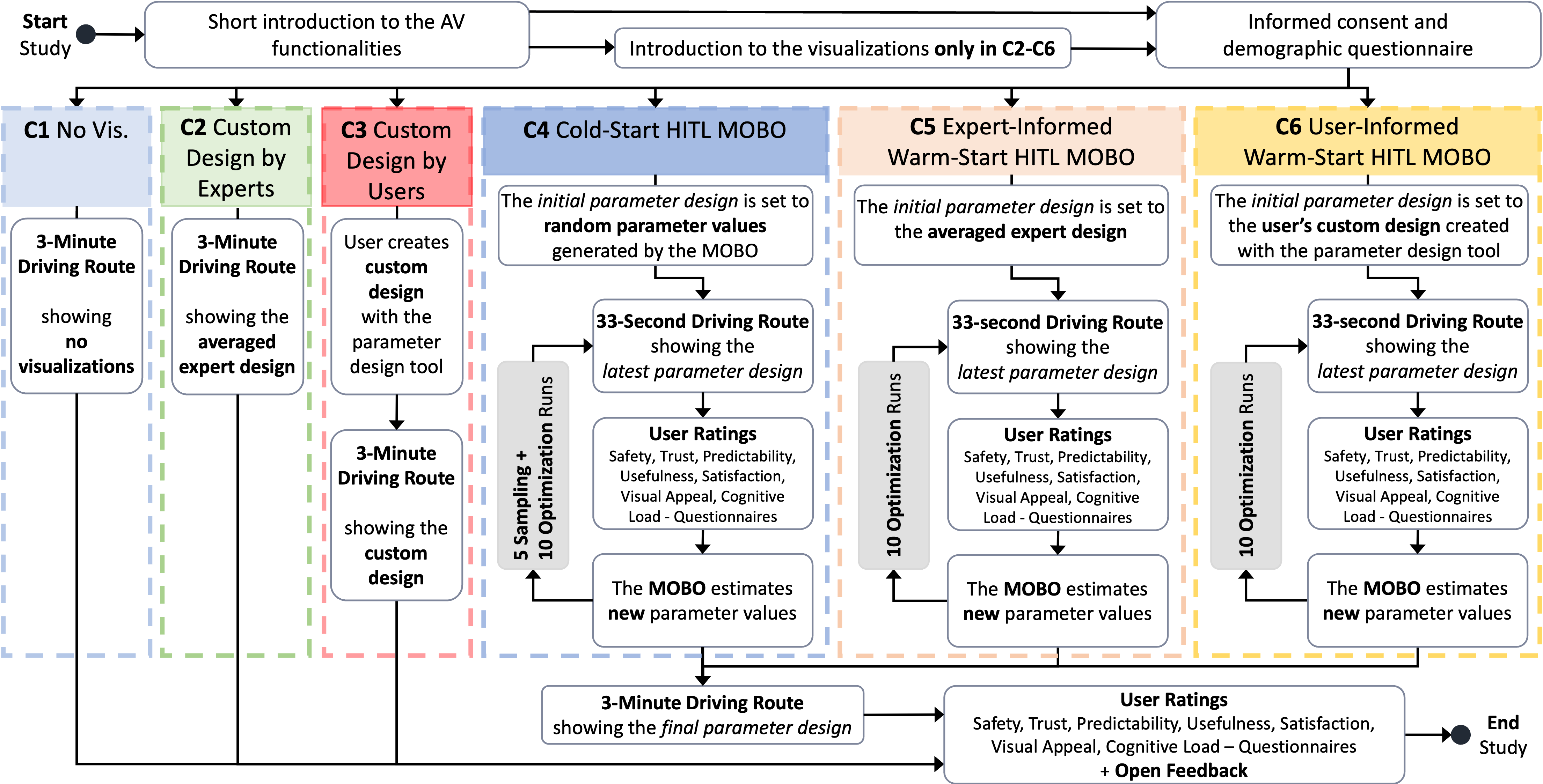}
   \caption{Study procedure of the six conditions C1-C6.}
   \label{fig:process}
    \Description{The figure shows the procedure for each of the six study groups C1-C6 using flow charts.}
\end{figure*}

Participants were distributed across the \textbf{six} conditions (see \autoref{fig:process}). Upon downloading the Unity driving simulation and the Bayesian optimizer (see Section~\ref{study-apparatus}), sessions began with a short introduction, informed consent, and a demographic questionnaire. The AV introduction was adapted from \citet{colley2022scene}.

For conditions C2-C6, participants were informed that the AV would display detected objects, their predicted actions, and the AV's planned maneuvers on its WSD (see \autoref{fig:intro}). Additionally, they received a brief overview of the visualizations with examples for semantic segmentation, pedestrian intention, trajectory, ego trajectory, CAD-covered area, occluded cars, and status HUD (see \autoref{fig:in-vehicle-visualizations-overview} a-g).
Despite the visualizations performing perfectly (i.e., always highlighting all relevant objects), participants were intentionally informed that the AV ''attempts to assess the situation,'' implying the possibility of errors during the ride. This introduced a sense of potential risk to establish subjective rating levels (e.g., trust, see \cite{lee2004trust}).

According to our optimizer setup (see Section \ref{method-hyperparam-setup}), the 33-second driving route (see \autoref{fig:driving_route}) was repeated 15 times (5 sampling and 10 optimization iterations) in C4 and 10 times in C5 and C6 (only optimization due to the available data in C5 and C6 which were used for the sampling phase). For the final user rating, we employed the 3-minute route to ensure the final designs were experienced in new situations. In non-MOBO conditions C2 and C3, participants only experienced the 3-minute route.
In C4-C6, they undertook multiple trips in the AV on the same route.
The total task time was 3 minutes for C1 and C2, up to 12 minutes for C3 (including up to 8 minutes of custom designing), 11.25 minutes for C4, and 8.5 minutes for C5 and C6.

We did not inform participants they were part of a HITL optimization process or detail how the optimizer applied their feedback to the design. In real-world scenarios, especially with in-vehicle interactions, we argue that users do not have deep knowledge about the system's operation but can still evaluate the quality of an experience. In total, the study lasted up to 50 minutes, and we integrated attention and comprehension checks following \href{https://researcher-help.prolific.co/hc/en-gb/articles/360009223553-Prolific-s-Attention-and-Comprehension-Check-Policy}{Prolific's guidelines}.
Participants could not intervene in the driving task as the visualizations are primarily intended to inform the user about AV functionalities.

\subsubsection{Subjective Ratings}\label{study-subjective-measures}
During the HITL optimization, participants rated the visualization designs via the subjective metrics defined in Section \ref{method-objectives} \textbf{after} each ride, with the possibility for textual feedback. For the C1-\textit{No Vis.} condition, we did not assess Acceptance (i.e., Usefulness and Satisfaction~\cite{van1997simple}) and Aesthetics as these would not make sense without any visualization.
After the session, in the final user rating, we measured the subjective metrics defined in Section \ref{method-objectives} and the design experience using adapted questions from \citet{chan2022bo}. On 7-point Likert scales (1=\textit{Strongly disagree} to 7=\textit{Strongly agree}), we queried about \textit{User Expectation Conformity}: "The final design matches my imagination.", \textit{Satisfaction}: "I'm pleased with the final design.", \textit{Confidence}: "I believe the design is optimal for me.", \textit{Agency}: "I felt in control of the design process." and \textit{Ownership}: "I feel the final design is mine." Regarding \textit{Interactivity}, participants also provided feedback on desired design control levels ("... Consider aspects where you desired more or less control over the design."). Participants could further elaborate with textual comments if they disagreed with the statements for expectation or satisfaction.

\subsubsection{Objective Measures}\label{study-objective-measures}
We recorded the Bayesian optimizer's performance metrics 
and the time taken in the Unity application for questionnaire responses (C2-C6) and custom design (C2 and C6). For RQ3, we embedded the webcam-based eye-tracker UnitEye \cite{10.1145/3670653.3670655} in the study application to track areas of interest (AOIs): pedestrian, vehicle, traffic sign, pedestrian intention icon, occluded car, CAD-covered area sphere, and vehicle status HUD. Participants calibrated the eye-tracker before the study so that we could monitor their focus and attention during the design and the AV ride on the 33-second and 3-minute driving routes. The eye-tracking data was not used as an objective function of the Bayesian optimizer.


\section{Results}\label{results}

\subsection{Quantitative Results}

\subsubsection{Data Analysis}
Before every statistical test, we checked the required assumptions (e.g., normality distribution). R in version 4.4.2 and RStudio in version 2024.09.0 were employed. All packages were up-to-date in December 2024.
We used the ARTool package by \citet{wobbrock2011art} for non-parametric data as the typical ANOVA is inappropriate with non-normally distributed data and Holm correction for post-hoc tests. The procedure is abbreviated with ART.
For the comparisons report in Section~\ref{sec:design}, we used \textbf{all Pareto front} values per MOBO condition per user. 
\autoref{fig:ps_and_cog} to \autoref{fig:aes} show only significant differences via bars using Dunn's test for post-hoc comparisons with Holm correction.  
The progression of the dependent variables \textbf{during} the MOBO iterations are shown in \autoref{fig:run_ps_cog}, \autoref{fig:run_trust_pred}, and \autoref{fig:run_acc_aes}.

The error bars represent bootstrap confidence intervals (i.e.,\\ mean\_cl\_boot). 
We refrain from reporting the Pareto front graphically due to (1) the high number of parameter value combinations---even with only five discrete levels for our nine continuous design parameters (four $s$ and five $\alpha$ values, see \autoref{tab:design_param}), there are approximately $5^9$ possible combinations---and (2) the resulting challenges in visualization due to the high number of dimensions ($p_1 \dots p_{16}$).

\subsubsection{Number of Applied Stopping Criterion}
The stopping criterion was met when all six design objectives received perfect scores in two consecutive iterations (see Section~\ref{method-stop-criterion}). This occurred for 12 of the 57 participants (21.05\%) interacting with a HITL MOBO variant: three in C4-Cold-Start, four in C5-Expert-Informed Warm-Start, and five in C6-User-Informed Warm-Start HITL MOBO. Achieving perfect (i.e., maximum/minimum) scores across multiple objectives is challenging, making the 21.05\% proportion notable. Besides, the trend lines in \autoref{fig:run_ps_cog} to \autoref{fig:run_acc_aes} and the high scores of the other participants suggest that additional iterations would result in more users reaching perfect scores. We interpret this as validation of our stopping criterion and the assumption that convergence occurs quickly.

\subsubsection{Design Performance}\label{sec:design}

\begin{figure*}[h!]
\centering
         \begin{subfigure}[b]{0.49\linewidth}
    \includegraphics[width=\linewidth]{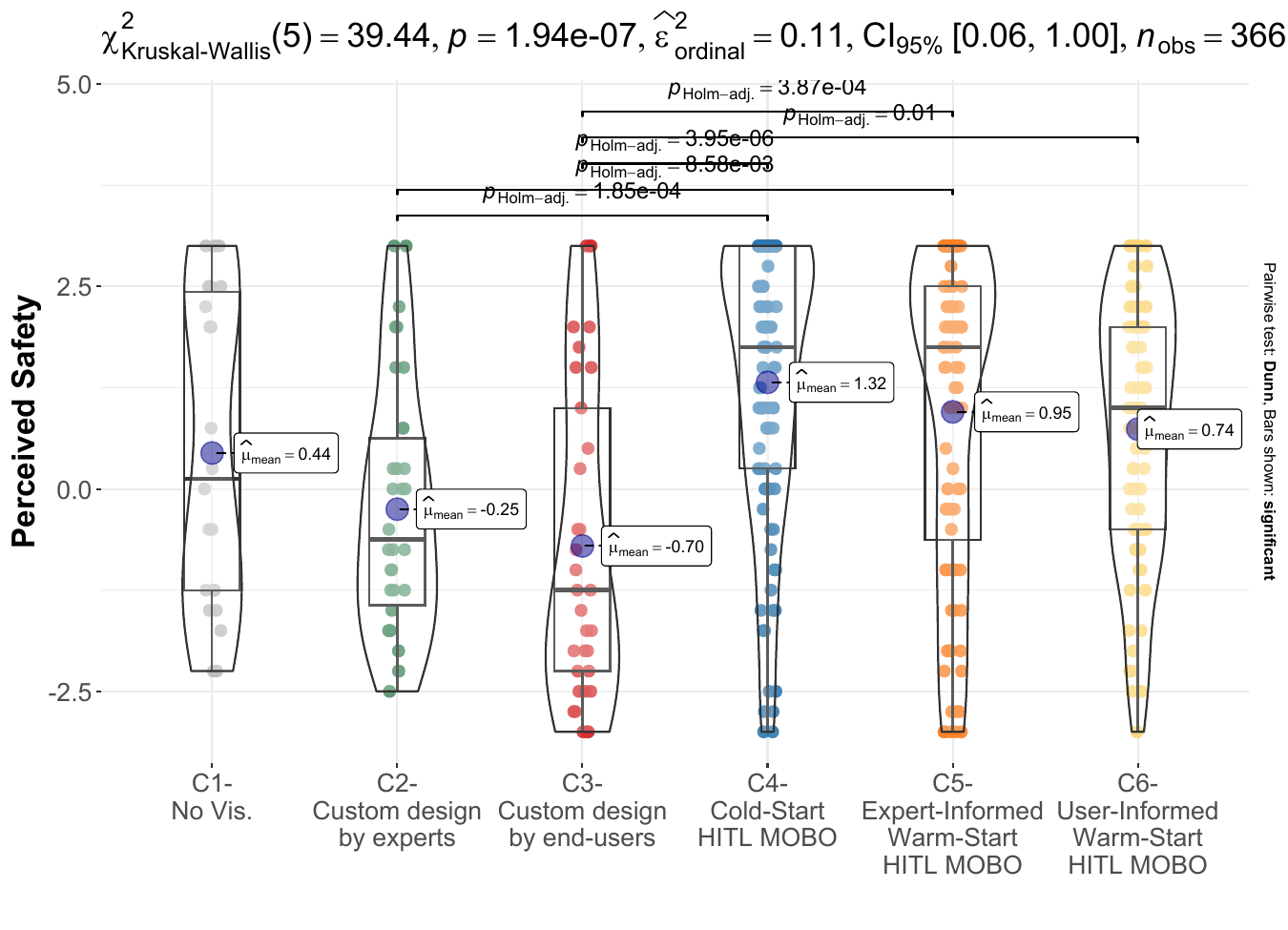}
   \caption{Statistical comparison of \textbf{perceived safety}.}
   \label{fig:ps}
    \Description{}
     \end{subfigure}
         \begin{subfigure}[b]{0.49\linewidth}
                          \includegraphics[width=\linewidth]{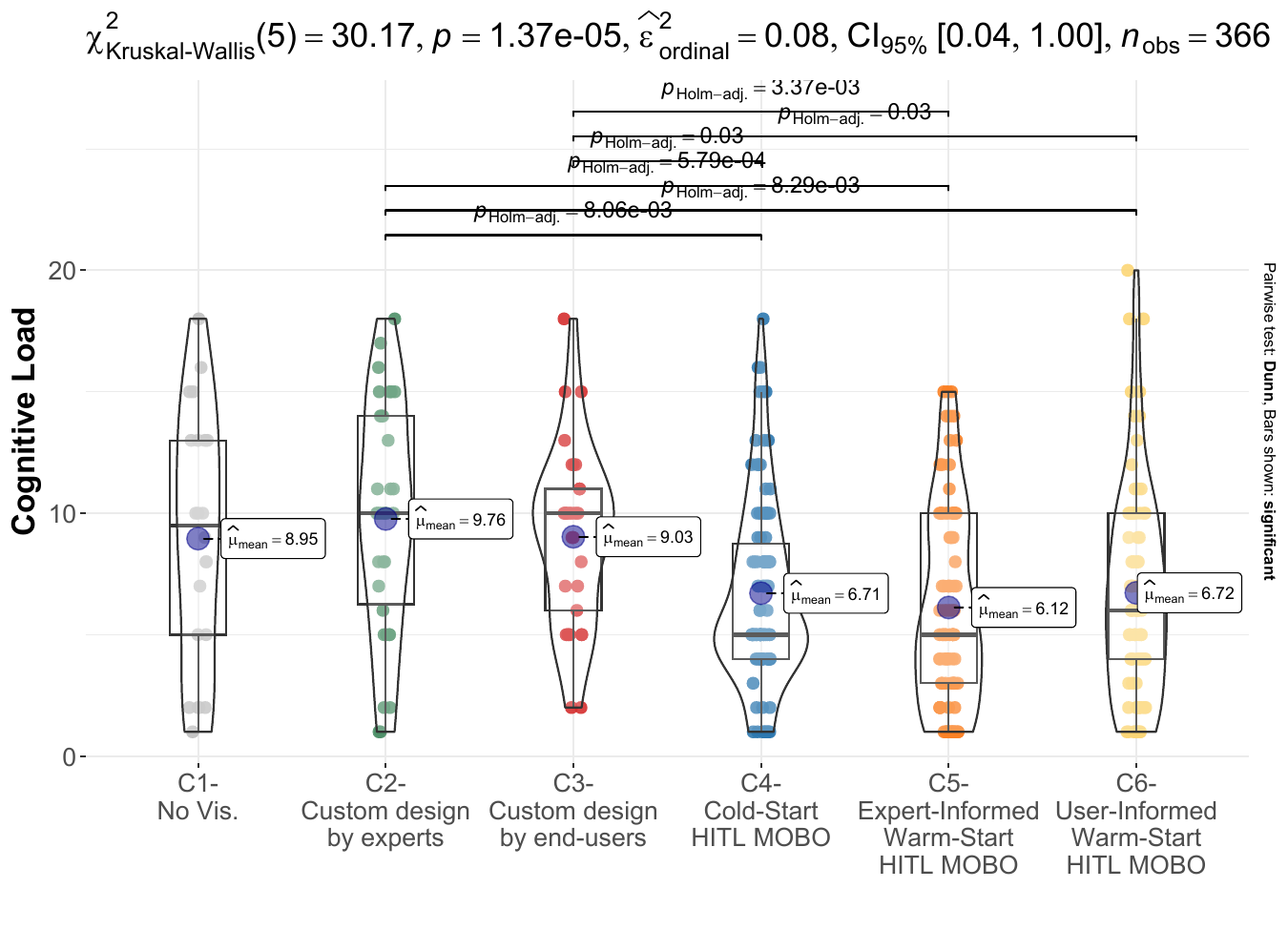}
   \caption{Statistical comparison of \textbf{cognitive load}.}
   \label{fig:cog}
    \Description{}
        \end{subfigure}
    \caption{Statistical comparison of perceived safety and cognitive load over the conditions.}\label{fig:ps_and_cog}
    \Description{Statistical comparison of perceived safety and cognitive load over the conditions.}
\end{figure*}

A Kruskal-Wallis rank sum test found a significant effect of \optimization on \perceivedSafetyScore (\chisq(5)=39.44, \pminor{0.001}, r=0.11; see \autoref{fig:ps}).

A post-hoc test found that C4-Cold-start HITL MOBO was significantly higher (\m{1.32}, \sd{1.72}) in terms of \perceivedSafetyScore compared to C2-Custom design by experts (\m{-0.25}, \sd{1.55}); \padjminor{0.001}), compared to C3-Custom design by end-users (\m{-0.70}, \sd{1.97}); \padjminor{0.001}).

A post-hoc test also found that C5-Expert-Informed Warm-Start HITL MOBO was significantly higher (\m{0.95}, \sd{1.99}) in terms of \perceivedSafetyScore compared to C2-Custom design by experts (\m{-0.25}, \sd{1.55}); \padj{0.009}) and compared to C3-Custom design by end-users (\m{-0.70}, \sd{1.97}); \padjminor{0.001}). 

 A post-hoc test finally also found that C6-User-Informed Warm-Start HITL MOBO was significantly higher (\m{0.74}, \sd{1.62}) in terms of \perceivedSafetyScore compared to C3-Custom design by end-users (\m{-0.70}, \sd{1.97}); \padj{0.014}). 



\noindent A Kruskal-Wallis rank sum test found a significant effect of \optimization on cognitive load (\chisq(5)=30.17, \pminor{0.001}, r=0.08; see \autoref{fig:cog}). A post-hoc test found that C2-Custom design by experts was significantly higher (\m{9.76}, \sd{4.72}) in terms of \MentalLoad compared to C4-Cold-start HITL MOBO (\m{6.71}, \sd{3.76}); \padj{0.008}), compared to C6-User-Informed Warm-Start HITL MOBO (\m{6.72}, \sd{4.31}); \padj{0.008}), and compared to C5-Expert-Informed Warm-Start HITL MOBO (\m{6.12}, \sd{4.05}); \padjminor{0.001}). 
A post-hoc test also found that C3-Custom design by end-users was significantly higher (\m{9.03}, \sd{3.64}) in terms of \MentalLoad compared to C4-Cold-start HITL MOBO (\m{6.71}, \sd{3.76}); \padj{0.032}), compared to C6-User-Informed Warm-Start HITL MOBO (\m{6.72}, \sd{4.31}); \padj{0.032}), and compared to C5-Expert-Informed Warm-Start HITL MOBO (\m{6.12}, \sd{4.05}); \padj{0.003}).


\begin{figure*}[h!]
\centering
             \begin{subfigure}[b]{0.49\linewidth}
             \includegraphics[width=\linewidth]{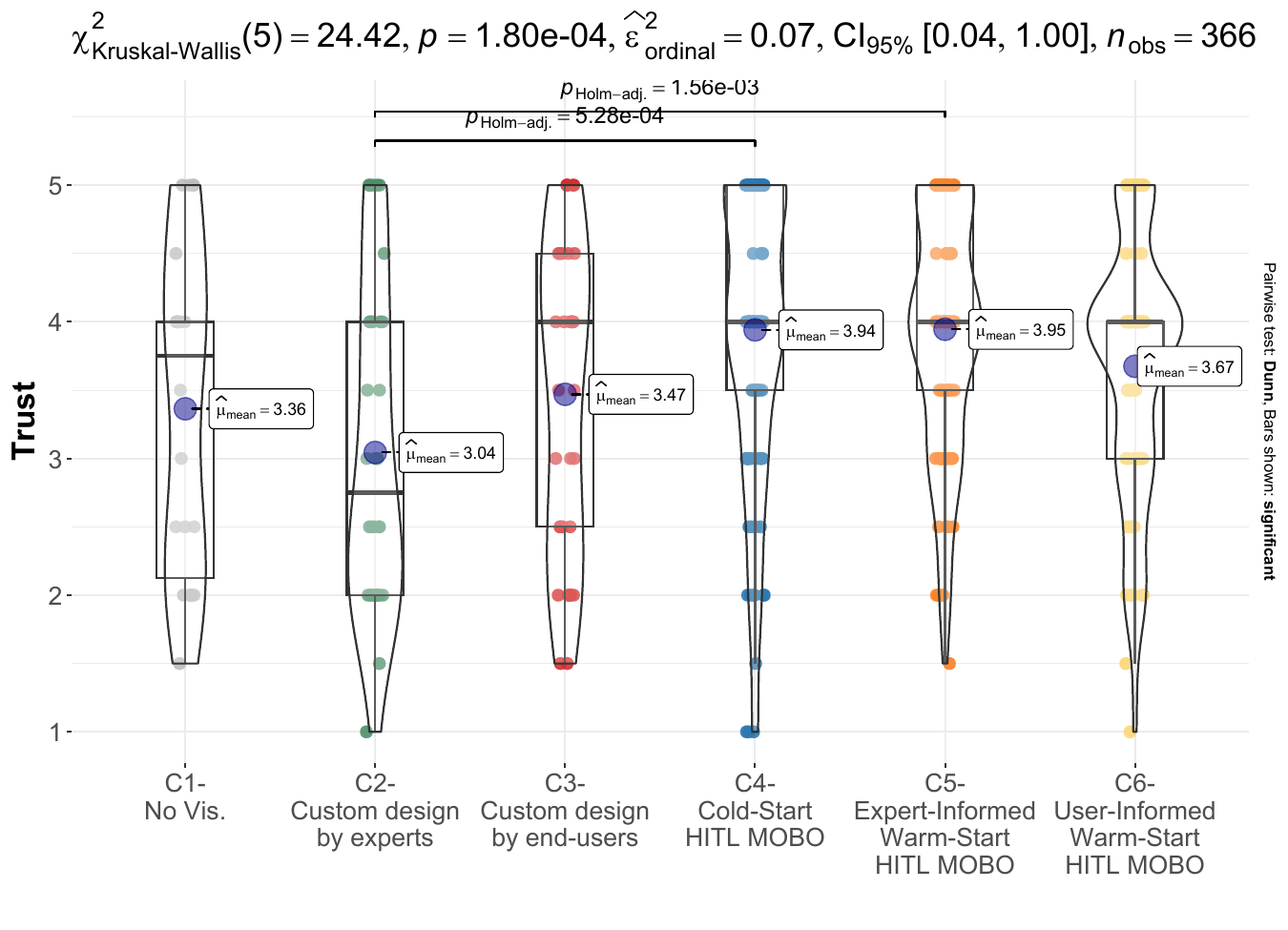}
   \caption{Statistical comparison of \textbf{trust}.}
   \label{fig:trust}
    \Description{}
              \end{subfigure}
         \begin{subfigure}[b]{0.49\linewidth}
    \includegraphics[width=\linewidth]{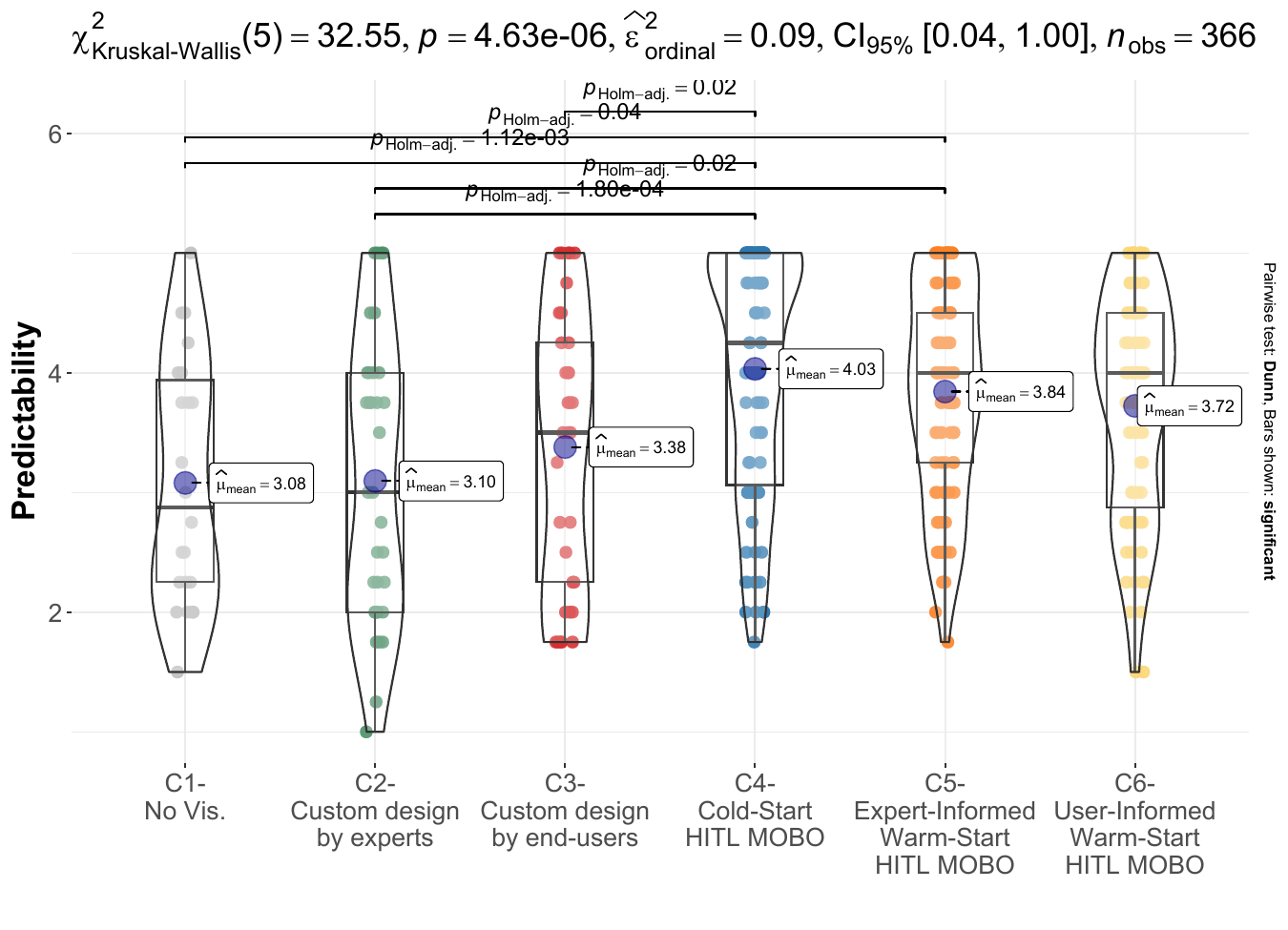}
   \caption{Statistical comparison of \textbf{predictability}.}
   \label{fig:pred}
    \Description{}
  \end{subfigure}
    \caption{Statistical comparison of trust and predictability over the conditions.}\label{fig:trust_and_pred}
        \Description{Statistical comparison of trust and predictability over the conditions.}
\end{figure*}



A Kruskal-Wallis rank sum test found a significant effect of \optimization on trust (\chisq(5)=24.42, \pminor{0.001}, r=0.07; see \autoref{fig:trust}). 
A post-hoc test found that C4-Cold-start HITL MOBO was significantly higher (\m{3.94}, \sd{1.07}) in terms of \trust compared to C2-Custom design by experts (\m{3.04}, \sd{1.17}); \padjminor{0.001}). A post-hoc test also found that C5-Expert-Informed Warm-Start HITL MOBO was significantly higher (\m{3.95}, \sd{0.89}) in terms of \trust compared to C2-Custom design by experts (\m{3.04}, \sd{1.17}); \padj{0.002}). 


A Kruskal-Wallis rank sum test found a significant effect of \optimization on predictability (\chisq(5)=32.55, \pminor{0.001}, r=0.09; see \autoref{fig:pred}). 
A post-hoc test found that C4-Cold-start HITL MOBO was significantly higher (\m{4.03}, \sd{1.00}) in terms of \predictability compared to C2-Custom design by experts (\m{3.10}, \sd{1.18}); \padjminor{0.001}), compared to C1-No Vis. (\m{3.08}, \sd{1.03}); \padj{0.001}), and compared to C3-Custom design by end-users (\m{3.38}, \sd{1.17}); \padj{0.021}). 
A post-hoc test also found that C5-Expert-Informed Warm-Start HITL MOBO was significantly higher (\m{3.84}, \sd{0.89}) in terms of \predictability compared to C2-Custom design by experts (\m{3.10}, \sd{1.18}); \padj{0.021}), compared to C1-No Vis. (\m{3.08}, \sd{1.03}); \padj{0.041})


\begin{figure*}[h!]
\centering
         \begin{subfigure}[b]{0.49\linewidth}
    \includegraphics[width=\linewidth]{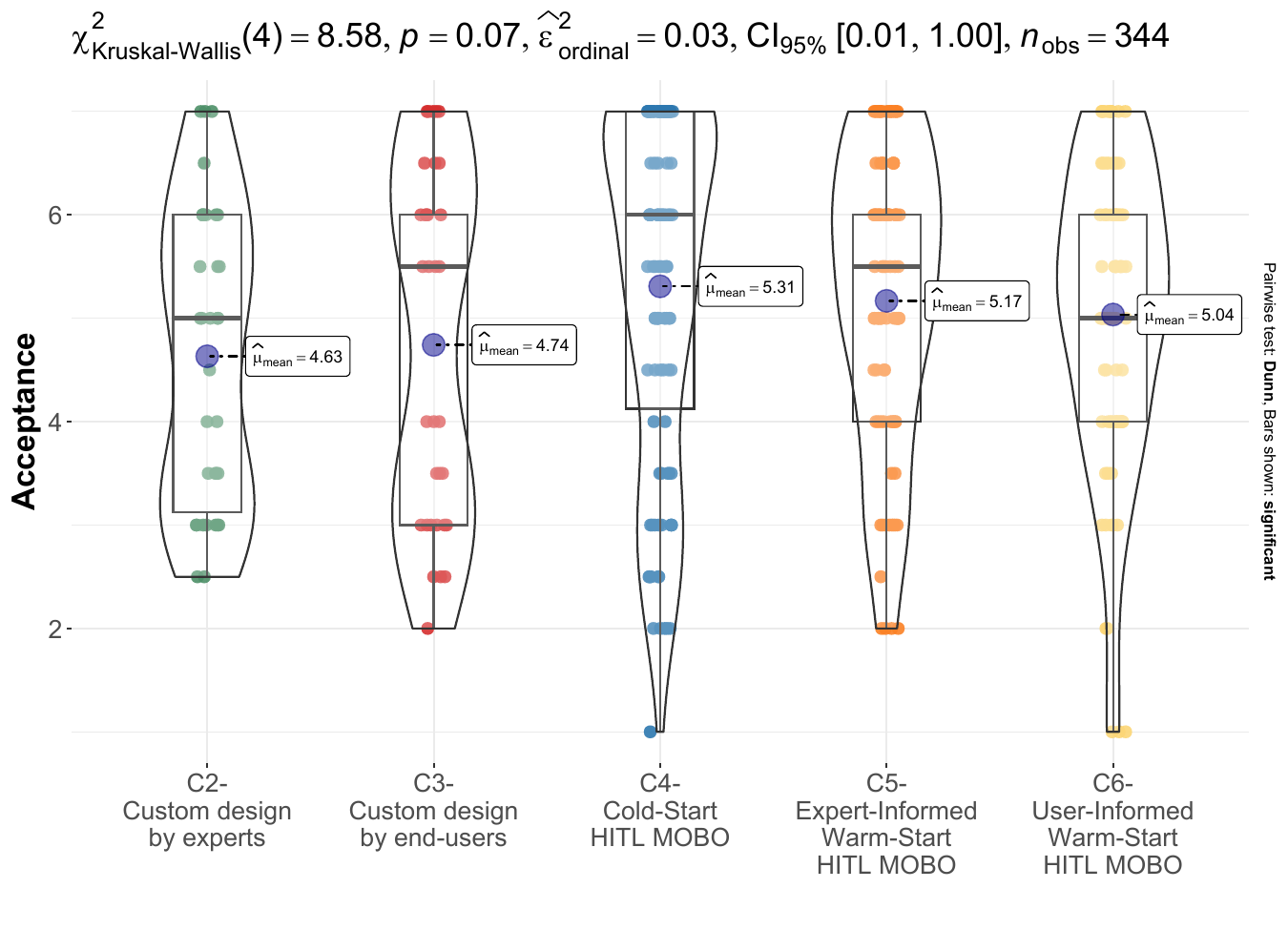}
       \caption{Statistical comparison of \textbf{acceptance}.}
   \label{fig:acc}
    \Description{}
      \end{subfigure}
             \begin{subfigure}[b]{0.49\linewidth}
                 \includegraphics[width=\linewidth]{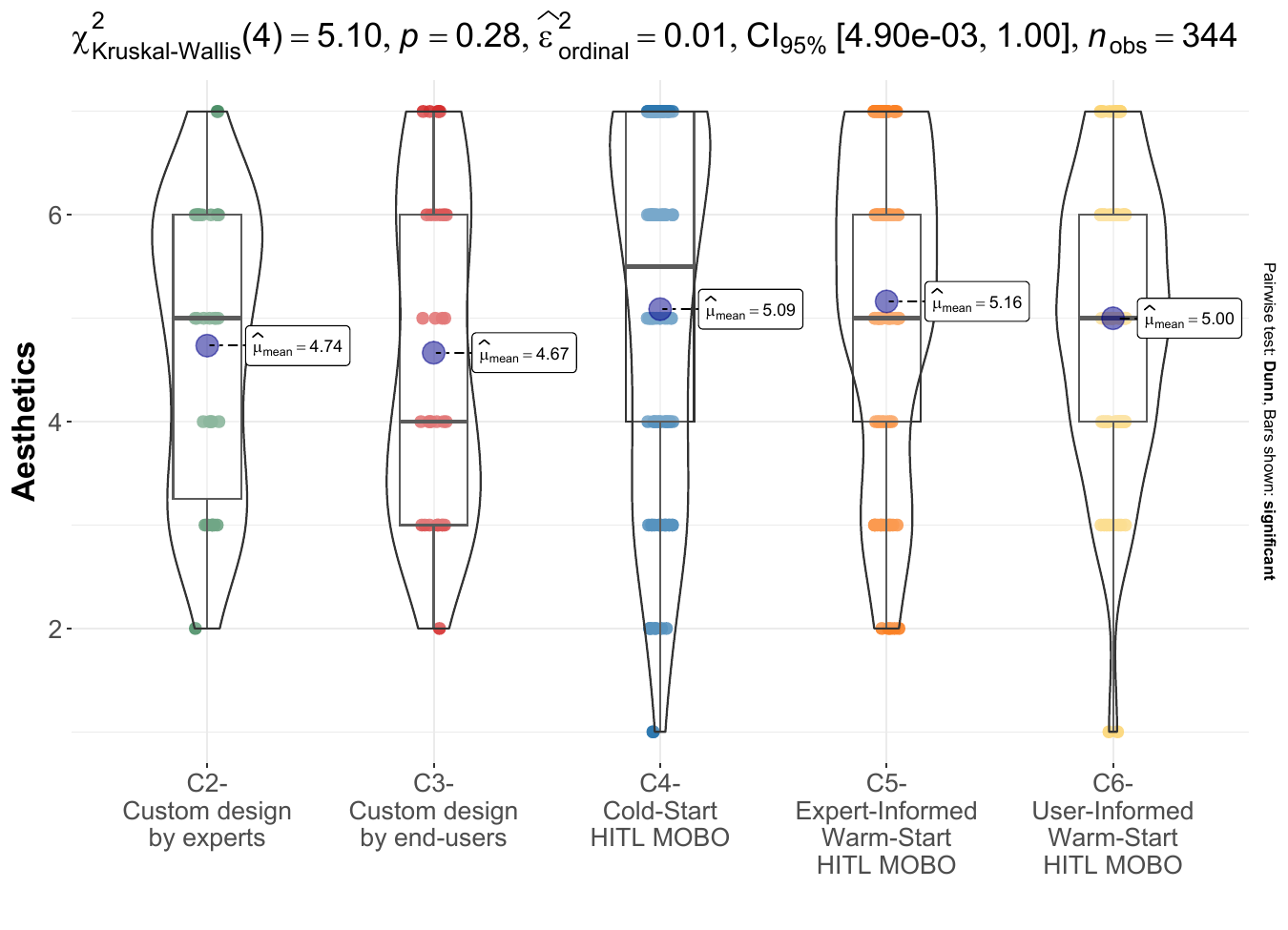}
   \caption{Statistical comparison of \textbf{aesthetics}.}
   \label{fig:aes}
    \Description{}
              \end{subfigure}
    \caption{Statistical comparison of acceptance and aesthetics over the conditions.  Data on acceptance and aesthetics was not collected for C1-\textit{No Vis.}}
    \Description{Statistical comparison of acceptance and aesthetics over the conditions.}
    \label{fig:acc-aes}
\end{figure*}

A Kruskal-Wallis rank sum test found no significant effects on acceptance (\chisq(4)=8.58, \p{0.073}, r=0.03; see \autoref{fig:acc}). 



A Kruskal-Wallis rank sum test found no significant effects on aesthetics (\chisq(4)=5.1, \p{0.277}, r=0.01); see \autoref{fig:aes}).
There was no visualization design to rate in C1-No Vis., so this condition is absent in \autoref{fig:acc-aes}.


\paragraph{Conclusion}
The HITL MOBO conditions (C4-C6) demonstrated significant improvements in \perceivedSafetyScore, \trust, and \predictability while reducing \MentalLoad compared to the non-MOBO conditions (C2, C3). However, no significant effects for acceptance and aesthetics were observed, providing \textbf{partial support for H1}. Moreover, \textbf{H2 is rejected}, as no significant differences were identified among the HITL MOBO conditions (C4-C6) for any design objective.


\subsubsection{Eye-Tracking Results}

The eye-tracking results indicated that participants were attentive to the study, with occasional divergent gazes away from the screen.
The ART found a significant main effect of \aoi (\F{4}{140}{13.04}, \pminor{0.001}) and of \GroupID on AOI fixation (\F{5}{35}{14.11}, \pminor{0.001}). The ART found a significant interaction effect of \aoi $\times$ \GroupID on AOI fixation  (\F{20}{140}{2.26}, \p{0.003}; see \autoref{fig:eye_gaze}).\\
For C6-User-Informed Warm-Start HITL MOBO, C2-Custom design by experts, and C3-Custom design by end-users, particular emphasis was placed on the speedometer. In C5-Expert-Informed Warm-Start HITL MOBO, emphasis was put on the car, which was also gazed upon comparatively frequently in the other conditions.

\subsection{Pareto Front Parameter Set}

\begin{figure*}[ht!]
\centering
    \includegraphics[width=\linewidth]{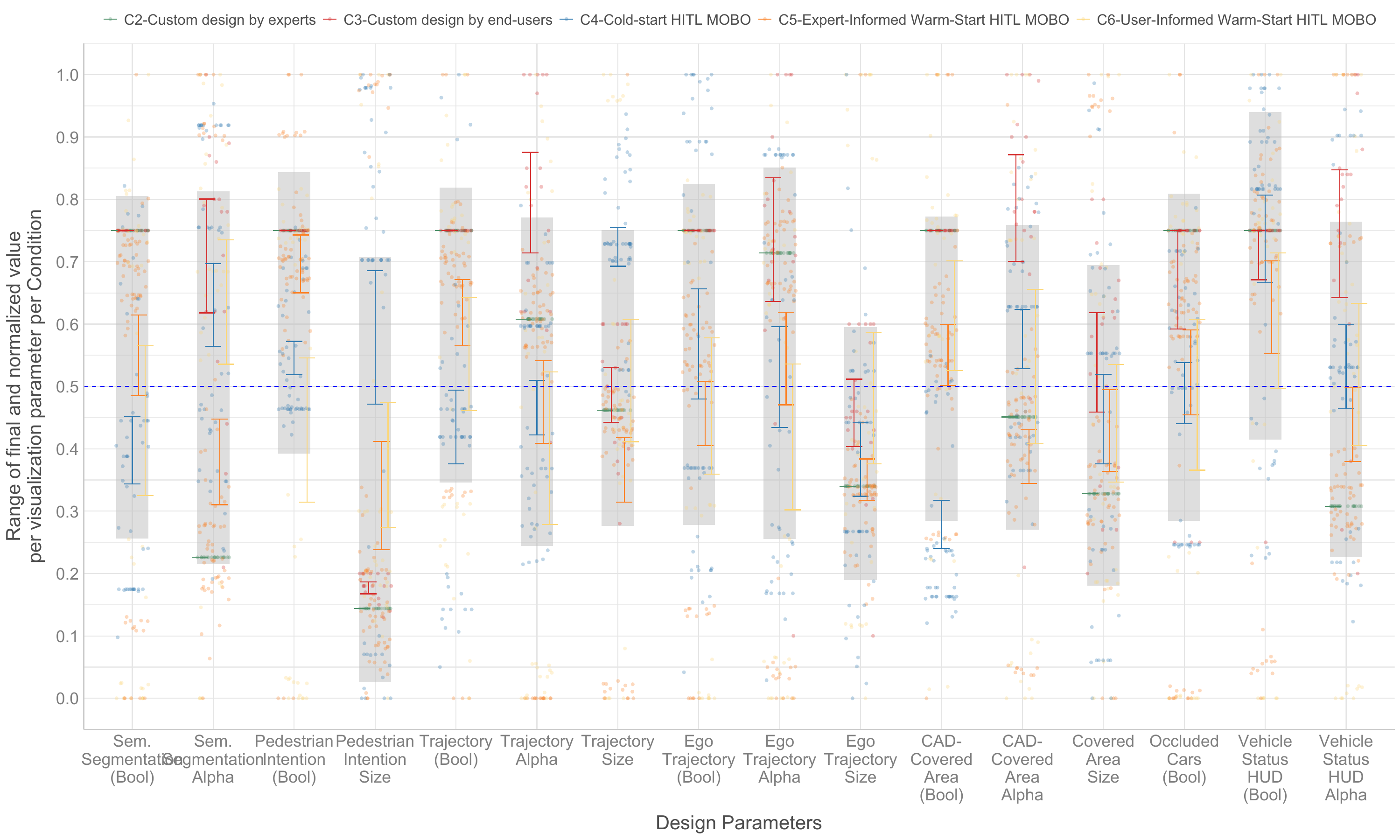}
   \caption{Final parameter set per condition. The jittered \textbf{Pareto front} values per participant are presented, normalized to $[0, 1]$. The gray rectangle shows one standard deviation from the mean of all values. The lines show the bootstrapped 95\% confidence intervals per condition. The x-axis shows the ordered parameters ($p_1$ to $p_{16}$) from left to right.}
   \label{fig:final_params}
    \Description{Linechart of the values of the final Pareto front design parameter set. }
\end{figure*}

\begin{figure*}[ht!]    
\centering
    \includegraphics[width=\linewidth]{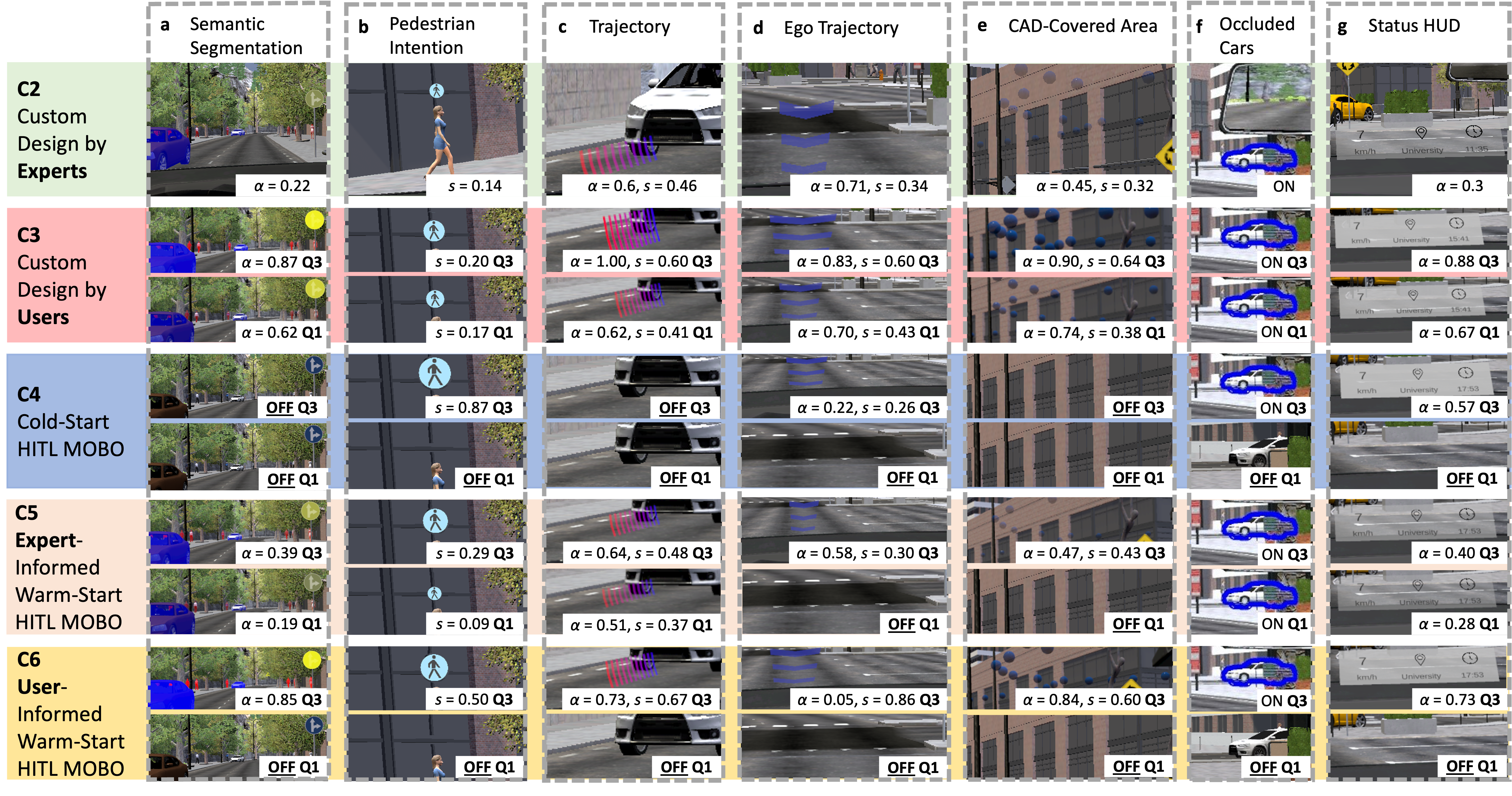}
   \caption{Visualization of parameter values from participants' Pareto front in conditions C3–C6 and the experts' \textit{standard} design in condition C2 (see \autoref{fig:final_params}). The parameters displayed are transparency ($\alpha$), size ($s$), and visibility ($v$). A visualization is \textit{OFF} if $v < 0.5$ and \textit{ON} if $v \geq 0.5$. C2 shows the mean parameter values from the experts' design. Conditions C3–C6 present participant parameter values at the interquartile range's lower (25th percentile, Q1) and upper (75th percentile, Q3) ends.} 
   \label{fig:final_params_visualized}
    \Description{The figure shows the mean values of the final parameter set for conditions C2-C6.}
\end{figure*}

\autoref{fig:final_params} shows the final parameter sets per condition (C2–C6), and \autoref{fig:final_params_visualized} visualizes these sets within the driving scene. Overall, in the HITL MOBO conditions (C4–C6), the bootstrapped 95\% confidence intervals suggest that many parameters were mostly turned off ($v < 0.5$), which contrasts C2 and rarely occurred in C3. Among the HITL MOBO conditions, C4 had the highest proportion of parameters turned off, likely due to the Cold-Start approach not incorporating prior knowledge, for example, informed by the experts' \textit{standard} visualization or users' custom design.

The HITL MOBO conditions exhibit wider intervals than the custom user design condition (C3), indicating a broad range of preferences not captured in the initial custom design phase. C3 showed less variation, possibly because participants found all visualizations initially relevant and enabled them out of curiosity. While these intervals provide insight into commonly preferred parameter ranges, individual differences persist. For instance, some participants activated the \textit{Vehicle Status HUD} while the majority did not, demonstrating that the intervals do not fully capture each user’s unique choices.

Regarding visualization transparency, C3 often had intervals above $\alpha = 0.7$, whereas other conditions hovered around $\alpha = 0.5$. This suggests that participants in C3 preferred clearer, more opaque visualizations when first encountering the scenario. Most conditions were similar in size parameters, but the \textit{Pedestrian Intention} visualization in C4–C6 showed larger intervals, pointing to greater preference diversity. This could mean that some participants preferred different parameter values as they became more familiar with the environment. Such variability underscores the value of approaches like \opticar, which can adapt to convergent and divergent user preferences.

\subsubsection{User Expectation Conformity, Satisfaction, Confidence, Agency, and Ownership}
A Kruskal-Wallis rank sum test found a significant effect of \optimization on user expectation conformity (\chisq(4)=10.62, \p{0.031}, r=0.11). However, post-hoc tests found no significant difference.

Kruskal-Wallis rank sum tests found no significant effects on Satisfaction (\chisq(4)=5.21, \p{0.266}, r=0.06),
Confidence (\chisq(4)=8.78, \p{0.067}, r=0.09), 
Agency (\chisq(3)=3.22, \p{0.358}, r=0.04), or
on Ownership (\chisq(3)=4.79, \p{0.188}, r=0.06).

\subsection{Qualitative Results}


After the final exposure to the visualizations, 22 participants provided open feedback on the expectation, satisfaction, and interactivity themes (see Section \ref{study-subjective-measures}). We analyzed this feedback using a structured two-phase process involving three authors. In the first phase, each author independently categorized the feedback into positive, negative, or suggestive sentiments across the three themes, summarizing key statements and insights.
The inter-rater agreement was assessed using Fleiss’ kappa, which yielded a value of 0.77, indicating substantial agreement \cite{landis1977measurement}.
In the second phase, the authors collaboratively reviewed and finalized which feedback to include, ensuring consistent sentiment assignment and resolving disagreements.

\subsubsection{Expectation} 
Analyzing the participants' expectations of the design reveals two distinct sentiments. Two positive comments emphasized comfort and safety, suggesting that the absence of excessive information made participants feel safer. One participant stated, \textit{"...for some reason, I felt more comfortable when I was not really seeing much of what the 'car' was 'thinking.'"} Another participant mentioned that the experience was less mentally demanding than expected, implying that an overly complicated interface can cause mental fatigue.

\noindent Conversely, negative feedback mainly revolved around design inconsistencies and inadequate visualization. Four participants showed concern when critical visual cues like the blue spheres (CAD-covered area visualization) disappeared, as noted, \textit{"...the blue bubbles showing the coverage area were also gone for some reason."} Others pointed out unnecessary or distracting design elements, such as \textit{"...the red jerky line and three blue lines..."} (referring to the trajectories). The overarching sentiment was a desire for more intuitive visualizations to understand the AV better.

\subsubsection{Satisfaction} 
Satisfaction levels varied among participants. Seven positive comments praised the effective visualization of pedestrians and vehicles (e.g.,  \textit{"I like the [...] way the car highlights everything, including the pedestrians and other vehicles}".
However, four participants voiced concerns about color coordination and the segmentation of certain objects, suggesting that some design elements could potentially confuse or distract the AV user. This sentiment is captured in the statement, \textit{"...some elements were too similar to each other in terms of color..."}. 

\subsubsection{Interactivity} 
Interactivity feedback illuminated participants' desire for more control and customization. Six positive remarks highlighted user satisfaction with the design optimization process, suggesting that personalized designs might increase user trust. A participant mentioned, \textit{"I liked having more control over the design. I feel like I would be helping a lot of people."}
The seven negative feedbacks highlighted issues with information overload (\textit{"I felt a little overwhelmed when the people, the vehicles, and the signs were all highlighted"}).  
Seven participants provided suggestions regarding visualization design enhancement and additional visualizations. Common suggestions included the addition of turn signals, clearer indications of the vehicle's route and intentions, and more dash notifications. One participant's comprehensive feedback, \textit{"...I wonder if there could be warning symbols and sounds when cars are braking..."} offers valuable insights into enhancing user trust through proactive system communications.

\paragraph{Conclusion}
Participants' feedback indicates increased engagement and alignment with user preferences through the HITL optimization process. The positive and negative feedback underscores the importance of intuitive design, clear visual cues, and user customization in building trust and ensuring user satisfaction.

\section{Discussion}

\subsection{Applicability of Bayesian Optimization on In-Vehicle Visualization Design}

The computational approach in \opticar effectively optimized the design objectives, as Cold-Start HITL MOBO (C4) yielded outcomes significantly superior (i.e., perceived safety, trust, predictability, usefulness, and satisfying) to the non-MOBO approaches (C2 and C3), which aligns with HCI literature~\cite{chan2022bo}. These findings partially support H1, as HITL MOBO conditions enhanced user ratings for most objectives but showed no significant improvements in acceptance or aesthetics. This suggests that querying user preferences in detail regarding visual design elements or cultural factors (see \cite{edelmann2021cross,lanzer2020designing}) in the HITL optimization process could help capture these subjective aspects more effectively.

The Warm-Start approaches (C5 and C6) also showcased significantly higher ratings than the non-MOBO approaches. However, these were less pronounced. Furthermore, no significant differences were observed between Cold-Start (C4) and Warm-Start (C5 and C6) conditions, leading to the rejection of H2, which hypothesized that the user-informed (C6) approach would outperform other HITL conditions. This might be attributed to imperfections in expert or user design data. Expert designs may not scale well to large sample sizes, and novice user custom designs can be flawed because user ratings may change after the initiation of the Bayesian optimizer as users get familiar with the situation. Thus, instead of enhancing the optimizer's exploration of the design space, the effectiveness of the Warm-Start approaches (C5 and C6) was slightly diminished compared to the Cold-Start variant.
Also, no significant differences between Cold-Start and Warm-Start HITL MOBO suggest that an averaged expert design could suffice as initial data for the optimizer in the AV functionality visualization design task.


In contrast to visualization designs resulting from traditional approaches (e.g., expert design, see \cite{colley2020effect, colley2022scene, colley2021effects, winter2019assessing, schneider2021explain}), MOBO can effectively identify more personalized designs. Traditional approaches require larger sample sizes and greater monetary incentives to achieve similar results while evaluating such broad design space in A/B testing. We show that MOBO is (at least) equivalent to traditional visualization design approaches in terms of its applicability by overcoming these testing constraints.


Although some participants reported qualitatively that they felt safer with less excessive information, this does not negate the need for feedback visualizations but requires designs that avoid interface clutter. This underscores a key challenge in AV visualization design, which is to augment information density to increase trust without amplifying the cognitive load~\cite{10.1145/3321335.3324947, colley2020effect, colley2022scene, colley2021effects, winter2019assessing, schneider2021explain}. Our findings suggest that the (Cold- and Warm-Start) HITL MOBO adeptly facilitates the optimization of all objectives, obviating the need for any trade-offs.


Safety is a key factor in the automotive sector. While HITL optimization is advantageous, it is critical to ensure that the resulting design alterations do not affect driving safety, if possible, also with future AVs. Although we consciously refrained from consulting some existing standards due to their orientation towards manual rather than automated driving, the \opticar method can be harmonized with such standards. For instance, we constrained certain design parameters, like the position of visualization elements, to avert overlapping UI elements (see Section~\ref{method-design-params}). Thus, future work should investigate the balance between customization and standardization in in-vehicle displays, particularly in scenarios where multiple end-users (e.g., in a shared AV) expect varying levels of information and functionality.


Creating scalable solutions for various vehicle models and contexts is challenging. However, our results demonstrate that the \opticar approach effectively generates visualization designs that receive high ratings in \textit{safety}, \textit{trust}, \textit{predictability}, \textit{usefulness}, \textit{satisfaction}, and \textit{aesthetics}. 
Additionally, as vehicles operate in dynamic environments \cite{bethge2021vemotion,jansen2023autovis}, the performance may improve with detailed driving context information (e.g., see \cite{bethge2021vemotion}). 
Thus, regular updates and re-optimizations are crucial, emphasizing the importance of long-term studies to capture evolving user needs and technology trends.

Besides, our findings indicate a convergence in end-user ratings towards a \textit{satisfying} level that does not necessarily represent the highest possible value of an objective (see Section \ref{method-objectives}). As the optimization progresses, for example, in long-term usage, the potential improvements in user trust, acceptance, or perceived safety from design changes might diminish. This diminishing return can lead to a saturation point, beyond which further optimization might not yield significant benefits or be cost-effective.

\subsection{Naturalistic User Reactions to Optimizer-Led Design Processes}

Integrating user recordings via webcam-based eye-tracking provides a unique perspective into user engagement and attention patterns during the HITL design process despite the noisiness of the data.
For instance, if users know that their design evaluation is being observed and validated, it is more likely that their feedback is genuine (see also Hawthorne effect \cite{oswald2014handling}). This can enhance the validity of the optimization process, ensuring that the ratings queried reflect the users' genuine experiences.

Analyzing the gaze patterns and AoIs revealed varying user attention across different conditions. However, we did not find a uniform pattern of attention across all participants or conditions. Regardless of the MOBO condition, other cars and the speedometer attracted a high fixation percentage. This indicates that such primary driving-related information is important to users. As AV technology becomes more prevalent, it is plausible that users' attention might transition to elements more pertinent to non-driving related activities (see \cite{pfleging2016investigating}).

Besides, eye-tracking data can be beneficial for refining the optimization process. We might assign a weighted importance metric by ascertaining which UI elements attract the most attention during classification.
This adaptive approach could ensure the optimization remains attuned to user preferences and needs.

\subsection{Empowering Non-Experts in Designing In-Vehicle Visualizations}

Typically, end-users provide feedback during A/B testing of in-vehicle visualization designs as part of a user-centered design process \cite{ISO9241_11_2018}. However, designers must interpret and iteratively integrate this feedback, a process that could benefit from direct incorporation through HITL design to meet individual needs and preferences better. Furthermore, while there is a clear demand for personalizing visualization designs \cite{10.1145/3409118.3475132}, current design approaches often restrict personalization within predefined limits set by designers. Additionally, implementing these manual personalizations can be challenging, frequently requiring users to navigate through setting menus.

\opticar presents a paradigm shift. By leveraging optimization-driven approaches, we demonstrate the potential to empower end-users to participate actively in the in-vehicle visualization design process. This democratizes the design and paves the way for optimized personalization, allowing for designs that simultaneously cater to multiple objectives. Such a framework can bridge the gap between designers' intents and users' preferences.
Another potential solution to better harness prior user knowledge could be to introduce a startup phase extending our investigation of Warm-Start HITL MOBO approaches (C5 and C6). This phase could involve querying users about their existing knowledge before their first drive and subsequently through brief follow-up questions. The optimization could be fast with a broad user base, and Pareto's optimal designs could be approached quickly. 

However, limitations exist, such as the explicit nature of the HITL process with subjective feedback and the iterative nature of multiple optimization iterations \cite{liao2023interaction, chan2022bo}. 


\subsection{Towards Implicit Design Optimizations of In-Vehicle Visualizations}

Our MOBO approach employs an explicit optimization loop that continuously requests feedback on users' subjective states. However, frequent queries can lead to user fatigue and reduce the accuracy of their responses \cite{chan2022bo}. Additionally, these requests may disrupt users' ongoing in-vehicle UI interactions.
Taking inspiration from \citet{koyama2022boassistant}, a shift from an explicit to an implicit optimization loop is conceivable. They leverage BO to learn design objectives by observing design exploration behaviors. The system then offers design suggestions based on these observations.
Likewise, we envisage an implicit MOBO process incorporated into vehicle use. Although this system may still operate within a loop, it would transition from explicit feedback (e.g., Likert scale ratings) to implicit end-user feedback. Such feedback could be derived from their interaction behaviors like the input error rate, physiological cues like heart rate, or psychological indicators like emotional states. Relevant approaches in this domain were discussed by \citet{10.1145/3546726} and Colley, Hartwig et al.~\cite{colley2024autotherm}, emphasizing non-intrusive feedback collection during vehicle use via interior cameras.


\subsection{Necessity of Optimization Explainability}


The explainability of automated systems, particularly in the context of feedback visualizations in AVs, is pivotal \cite{colley2020effect, colley2022scene, colley2021effects, winter2019assessing, schneider2021explain, koo2015did}. Trust in automated systems, an essential component for user acceptance, is often intertwined with the user's comprehension of the system's behavior \cite{colley2020effect, colley2022scene, colley2021effects}. Thus, users may require a sound understanding of the mechanisms underpinning the adaptive nature of the UI enabled by \opticar to foster trust.

However, the challenge lies in communicating complex optimization algorithms, like MOBO, both transparent and comprehensible to non-expert users. Our study revealed that user agency and satisfaction remained high across all conditions (MOBO and non-MOBO). This suggests that participants also were content with the overall optimization process. Still, it is unclear to what extent they have understood it. It is noteworthy, though, to discern which specific facets of the optimization process contributed to the sense of satisfaction and perceived agency. Our results align with and extend the findings by \citet{chan2022bo}. Future research endeavors should delve deeper into understanding the nuances of users and how they shape their interactions with automated systems.

\subsection{Limitations and Future Work}

Our work has been instrumental in assessing the application of MOBO to improve the user experience of AV functionality visualizations. However, limitations exist.
The choice of algorithm, including the acquisition function and other parameters, might have influenced the study outcomes. As with any optimization approach, the selected hyperparameters can significantly impact the results. Along with this, the participant selection might introduce biases, which could affect the generalizability of our findings.
Besides, a limitation was using a webcam-based eye-tracking method, which has inherent inaccuracies~\cite{10.1145/3517031.3529615}.
Nonetheless, we argue that the inaccuracies are a necessary trade-off because the technique offered novel insights into users' reactions to the HITL process and the resulting designs that are otherwise infeasible for an online study.

The study's 33-second observation period per MOBO iteration is brief, which may limit the depth of understanding about the AV's functionalities and limits. However, previous work also employed short durations (e.g., one minute~\cite{colley2022scene}). Another limitation is that the 33-second route cannot cover every possible driving scenario. However, the optimal parameter values likely vary depending on the scenario. We argue that our results can still be generalized to most urban scenarios with cars, pedestrians, road crossings, and roundabouts. This is also shown by the final 3-minute route, where the optimized designs were still effective even though the scenario changed (e.g., an intersection was introduced). However, future work should investigate other environments, such as motorways, and consider different traffic dynamics, such as pedestrian densities.

In the MOBO conditions, participants had more exposure to the 33-second route than those in non-MOBO conditions. This repeated exposure, an inherent aspect in iterative HITL processes, likely increased familiarity with the scenario. However, we argue this was mitigated by showing the unknown 3-minute route for the final assessment across all conditions and the frequently changing visualization designs during MOBO iterations. Yet, future work should explore increased scenario exposure in non-MOBO conditions to investigate this further.

While the online study environment with driving videos on a computer screen provides a controlled environment with high internal validity, it may have implications for generalizability in real-world driving scenarios. Our results are generalizable despite the wider field of view and possible physical consequences in the real world. In the real world, end-users would also focus on the scene directly in front of the AV and be less interested in what happens behind it. Even without real-world consequences, participants can realistically reflect their subjective perceptions in driving simulations~\cite{wang2010validity}. 
Besides, considering the complexity of real-world driving scenarios and the potential traffic dangers our prototype system could have introduced, a simulated environment was deemed suitable for this initial exploration of HITL MOBO in the automotive domain. Yet, future studies should prioritize the execution in real vehicles under dynamic driving conditions, for example, using approaches like XR-OOM~\cite{xroom} or PassengXR~\cite{PassengXR}. If technically impractical, a vehicle motion simulator can be used (e.g., SwiVR Car-Seat~\cite{colley2021swivr}).

Interface clutter is a recognized issue~\cite{colley2020effect, colley2022scene, colley2021effects}. An interface attempting to display all potential visualization parameters could become overwhelming, potentially distracting users or obscuring essential information. In our study, while we aimed for comprehensive visualization, the risk of cluttering the interface remained.

Cultural differences influence user perceptions of AVs \cite{edelmann2021cross,lanzer2020designing} (e.g., acceptance of AV maneuvers \cite{edelmann2021cross}), also likely affecting their perception of AV functionality visualizations. Our study was limited to one culture. However, the \opticar approach using HITL optimization could accommodate diverse user needs stemming from cultural context. To assess its effectiveness, future research should evaluate \opticar with users from different cultures.



\section{Conclusion}
This work employed HITL MOBO to navigate the design space of AV functionality visualizations on AR WSDs. An online study with N=117 participants helped validate the MOBO-driven approach, confirming its efficacy in optimizing visualizations for AVs. While statistical significance was not reached compared to No Visualization (C1) besides for predictability, the ratings were \textbf{always} better for C4-Cold-Start HITL MOBO.
Besides, this Cold-Start and Warm-Start optimization significantly improved perceived safety, cognitive load, and trust. Acceptance and aesthetics did not differ over the conditions.
Consequently, \opticar provides a pathway for creating personalized in-vehicle visualization designs that improve end-user experiences and decrease development time and expenses.

\section*{Open Science}
We make the Bayesian optimizer (see \url{https://github.com/Pascal-Jansen/Bayesian-Optimization-for-Unity}), the Unity application upon request, and the collected (anonymized) data (see \url{https://github.com/M-Colley/opticarvis-data}) available. The Unity project supports novel application scenarios by including easily adaptable settings regarding MOBO, server-client infrastructure for the online study, webcam-based eye tracking~\cite{10.1145/3670653.3670655}, and AV driving behavior.

\begin{acks}
We thank all study participants. Additionally, we thank the 6th Summer School on Computational Interaction\footnote{\url{https://cixschool2022.cs.uni-saarland.de/}} for providing significant insights into computational optimization.
This research was supported by a German Academic Exchange Service (DAAD) fellowship and conducted in the context of the DFG project "Computational Optimization of In-Vehicle User Interface Design for Maximized Usability, Safety, and Trust."
\end{acks}

\bibliographystyle{ACM-Reference-Format}
\bibliography{sample-base}

\appendix

\section{Expert Study to Inform the Standard Visualization Design}\label{app:expert}

\begin{table*}[!h]
    \centering
    \scriptsize
    \caption{Results of the expert study regarding the 16 design parameters normalized to a $[0, 1]$ range. Experts (E1 - E8) used the custom parameter design tool to manually select parameters they deemed fitting for the given driving scenario.}
    \resizebox{\textwidth}{!}{%
    \begin{tabular}{p{3.8cm}p{0.8cm}p{0.8cm}p{0.8cm}p{0.8cm}p{0.8cm}p{0.8cm}p{0.8cm}p{0.8cm}|p{0.8cm}p{0.8cm}}
    \hline
    \textbf{Design Parameter} & \textbf{E1} & \textbf{E2} & \textbf{E3} & \textbf{E4} & \textbf{E5} & \textbf{E6} & \textbf{E7} & \textbf{E8} & \textbf{Mean}& \textbf{SD} \\ \hline

    \textbf{$p_1$: Sem. Segmentation} & 0.75 & 0.75 & 0.75 & 0.75 & 0.75 & 0.75 & 0.25 & 0.75 & 0.69 & 0.18 \\
    \textbf{$p_2$: Sem. Segmentation Alpha} & 0.10 & 0.37 & 0.10 & 0.10 & 0.19 & 0.10 & 0.36 & 0.49 & 0.23 & 0.16 \\ \hdashline

    \textbf{$p_3$: Pedestrian Intention} & 0.75 & 0.75 & 0.25 & 0.75 & 0.75 & 0.75 & 0.75 & 0.75 & 0.69 & 0.18 \\
    \textbf{$p_4$: Pedestrian Intention Size} & 0.19 & 0.20 & 0.14 & 0.10 & 0.11 & 0.15 & 0.10 & 0.17 & 0.15 & 0.04 \\ \hdashline
    
    \textbf{$p_5$: Trajectory} & 0.75 & 0.75 & 0.75 & 0.75 & 0.75 & 0.25 & 0.75 & 0.75& 0.69 & 0.18 \\
    \textbf{$p_6$: Trajectory Alpha} & 0.63 & 1.00 & 0.45 & 0.41 & 0.31 & 0.54 & 0.52 & 1.00 & 0.61 & 0.26 \\
    \textbf{$p_7$: Trajectory Size} & 0.52 & 0.59 & 0.60 & 0.51 & 0.17 & 0.34 & 0.46 & 0.51 & 0.46 & 0.14 \\ \hdashline
    
    \textbf{$p_8$: Ego Trajectory} & 0.75 & 0.25 & 0.75 & 0.75 & 0.75 & 0.25 & 0.25 & 0.75 & 0.56 & 0.26 \\
    \textbf{$p_{9}$: Ego Trajectory Alpha} & 0.72 & 0.87 & 0.71 & 0.53 & 0.76 & 0.55 & 0.97 & 0.60 & 0.71 & 0.15 \\
    \textbf{$p_{10}$: Ego Trajectory Size} & 0.41 & 0.52 & 0.17 & 0.28 & 0.26 & 0.35 & 0.35 & 0.38 & 0.34 & 0.10 \\ \hdashline

    \textbf{$p_{11}$: CAD-Covered Area} & 0.75 & 0.75 & 0.25 & 0.25 & 0.75 & 0.75 & 0.25 & 0.75 & 0.56 & 0.26 \\
    \textbf{$p_{12}$: CAD-Covered Area Alpha} & 1.00 & 0.94 & 0.10& 0.10 & 0.21 & 0.78 & 0.10 & 0.38 & 0.45 & 0.39 \\
    \textbf{$p_{13}$: CAD-Covered Area Size} & 0.66 & 0.36 & 0.20 & 0.20 & 0.26 & 0.32 & 0.20 & 0.42 & 0.33 & 0.16 \\ \hdashline

    \textbf{$p_{14}$: Occluded Cars} & 0.75 & 0.75 & 0.25 & 0.75 & 0.75 & 0.75 & 0.75 & 0.25 & 0.63 & 0.23 \\ \hdashline
    
    \textbf{$p_{15}$: Vehicle Status HUD} & 0.75 & 0.75 & 0.75 & 0.75 & 0.75 & 0.75 & 0.75 & 0.75 & 0.75 & 0.00 \\
    \textbf{$p_{16}$: Vehicle Status HUD Alpha} & 0.10& 0.13 & 0.54 & 0.10 & 0.20 & 0.72 & 0.36 & 0.32 & 0.31 & 0.23 \\ \hline    
    \end{tabular}
    \label{tab:expert}
    }
\end{table*}

\section{Procedure}\label{app:study-procedure}

AV functionality introduction:

\begin{quote}
\textit{You will see a video of a driving session in an automated vehicle. The vehicle takes over lateral and longitudinal control (braking, accelerating, steering). The vehicle attempts to assess the situation and determine the intent of nearby pedestrians and cars. While watching the video, you are supposed to imagine sitting in such an automated vehicle, following the entire journey attentively, and then assessing it.}
\end{quote}

Introduction to the trips for C4-C6.
\begin{quote}
\textit{After each trip, please rate the visualization. The vehicle will adapt its design based on your feedback. This continues until the vehicle finalizes a design, and you'll take one more ride with that selected design.}
\end{quote}

\section{Results}

\subsection{Objective Values over Iterations}

\begin{figure*}[ht!]
\centering
         \begin{subfigure}[b]{0.49\linewidth}
    \includegraphics[width=\linewidth]{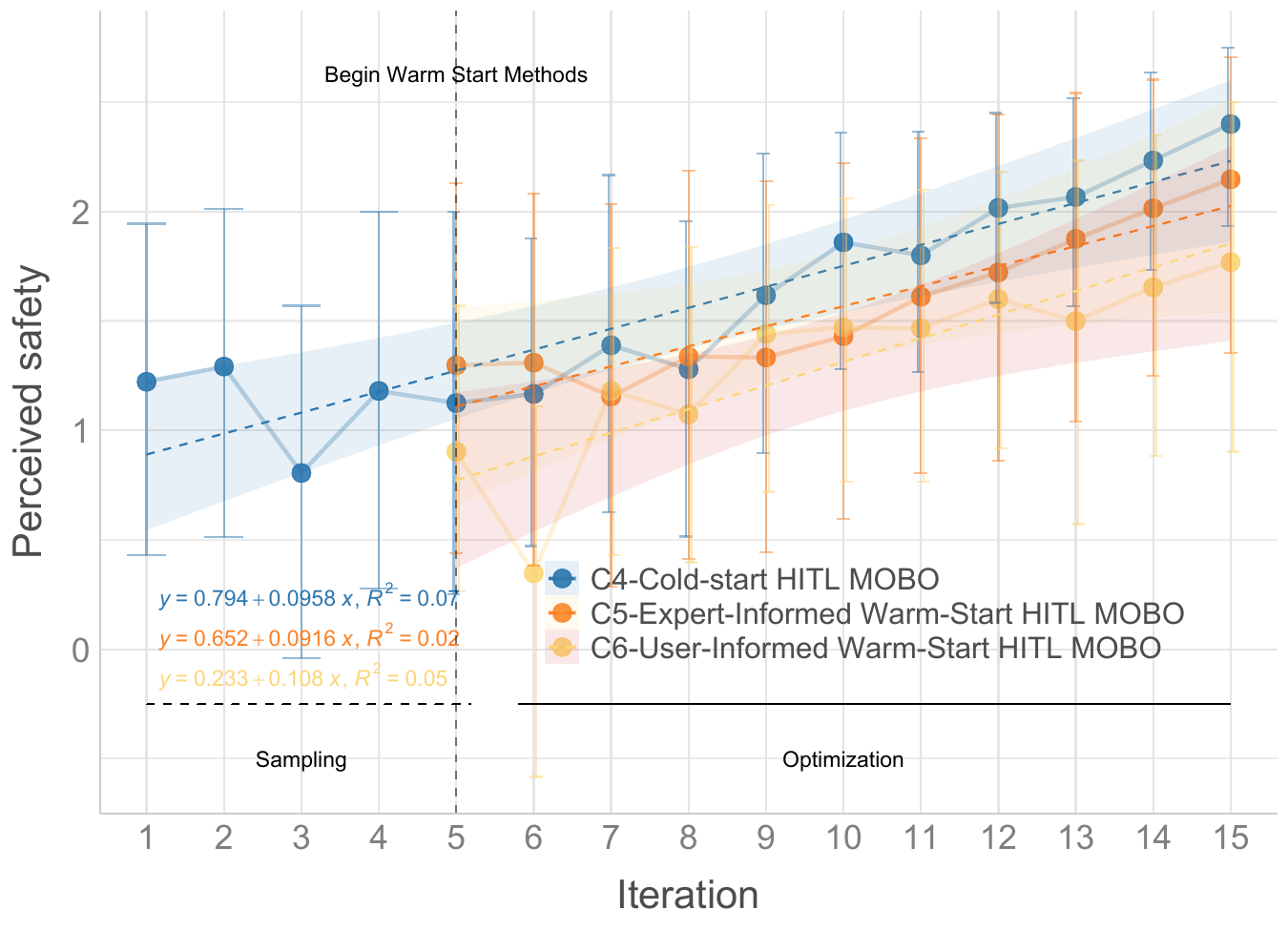}
   \caption{Progression of \textbf{perceived safety} over the MOBO iterations.}
   \label{fig:runs_ps}
    \Description{}
     \end{subfigure}
         \begin{subfigure}[b]{0.49\linewidth}
                          \includegraphics[width=\linewidth]{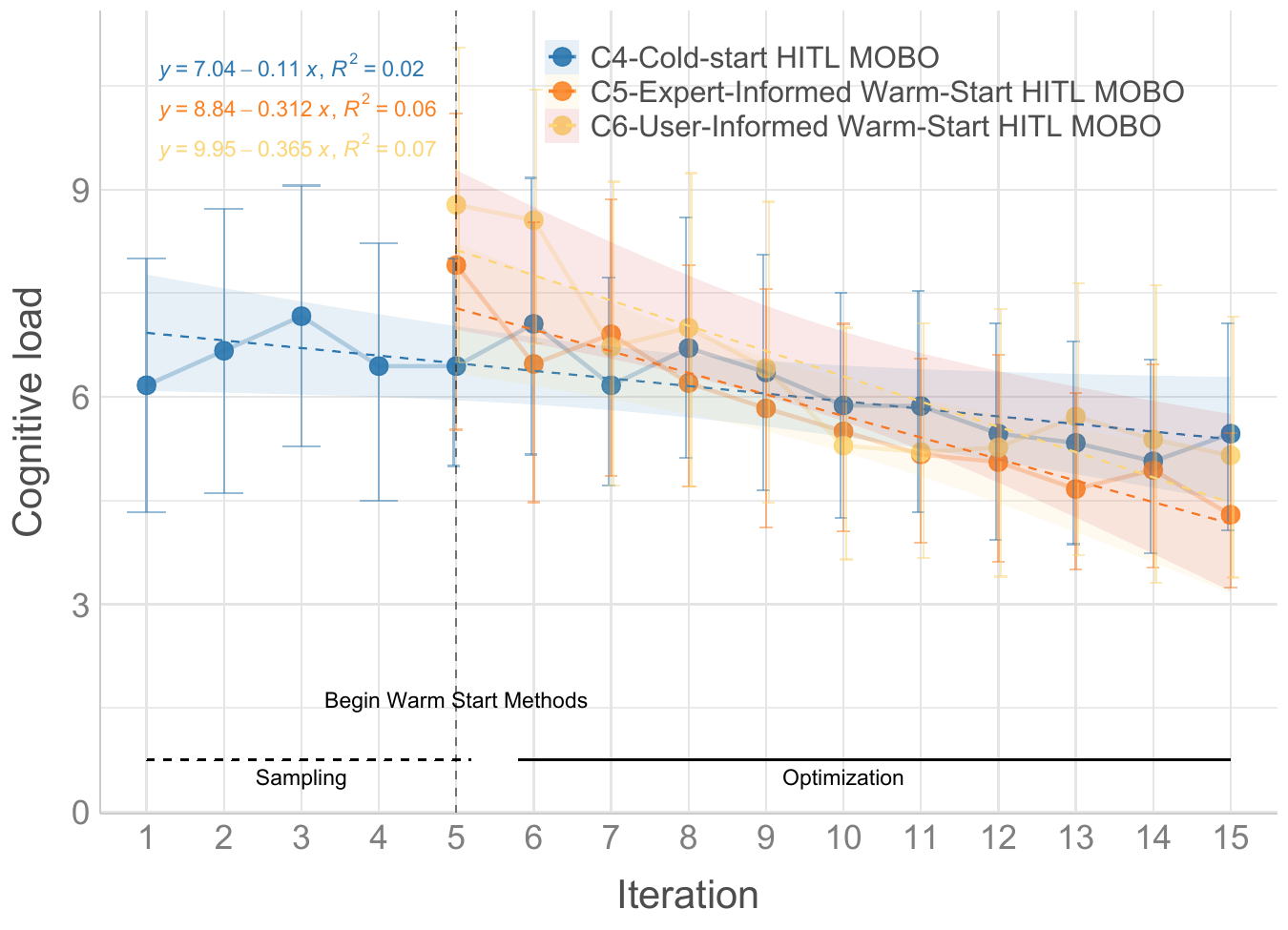}
   \caption{Progression of \textbf{cognitive load} over the MOBO iterations.}
   \label{fig:runs_cog}
    \Description{}
        \end{subfigure}
    \caption{Value progression of perceived safety and cognitive load. The Warm-Start conditions had \textbf{no} sampling phase (i.e., iteration five was their first iteration) as they were initialized by the averaged expert design (in C5) or a custom user design (in C6).}
    \Description{Progression of perceived safety (going steadily upward) and cognitive load (going steadily downward) over the MOBO iterations.}
    \label{fig:run_ps_cog}
\end{figure*}


\begin{figure*}[ht!]
\centering
             \begin{subfigure}[b]{0.49\linewidth}
             \includegraphics[width=\linewidth]{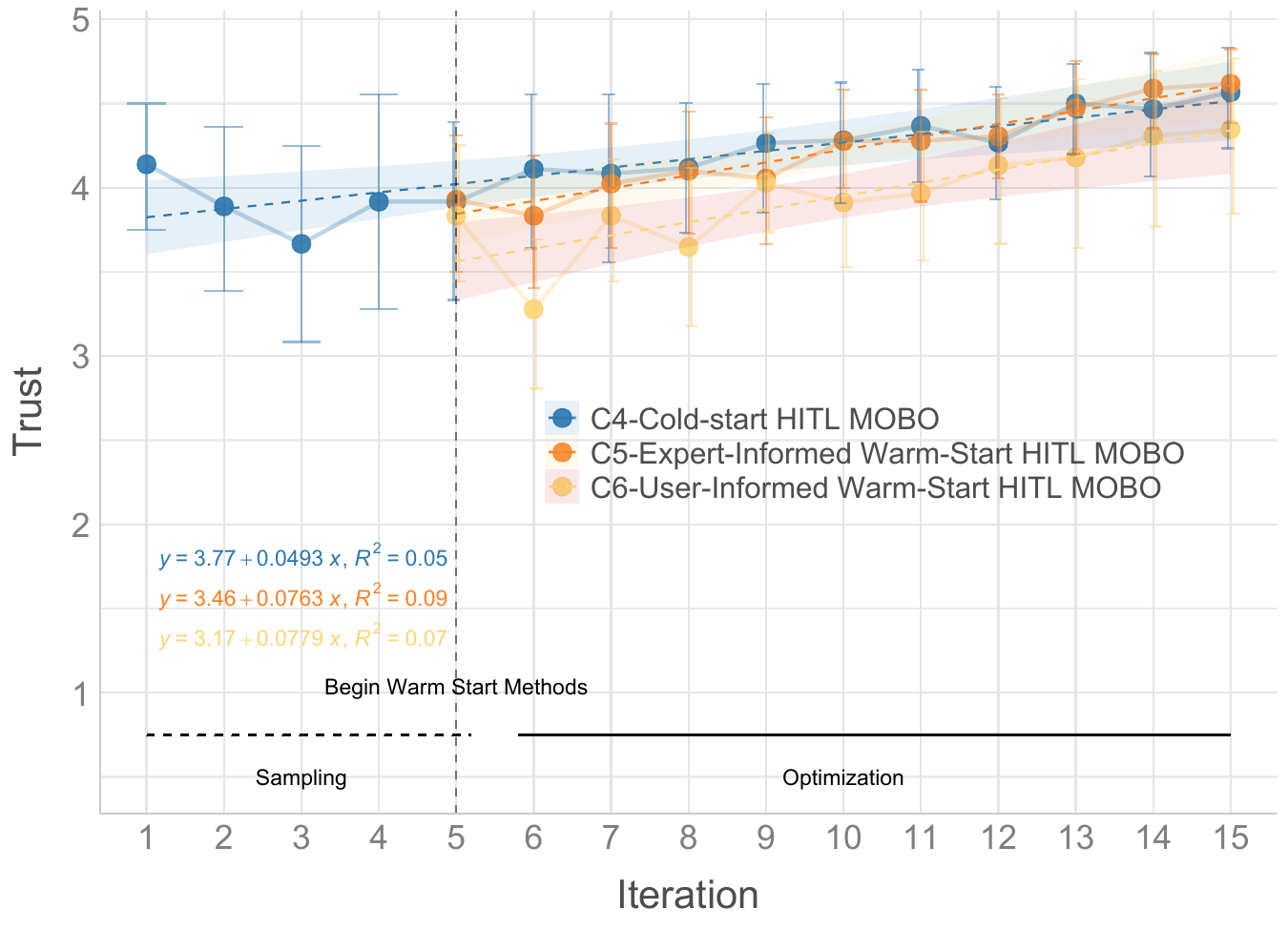}
   \caption{Progression of \textbf{trust} values over the MOBO iterations.}
   \label{fig:runs_trust}
    \Description{}
              \end{subfigure}
         \begin{subfigure}[b]{0.49\linewidth}
    \includegraphics[width=\linewidth]{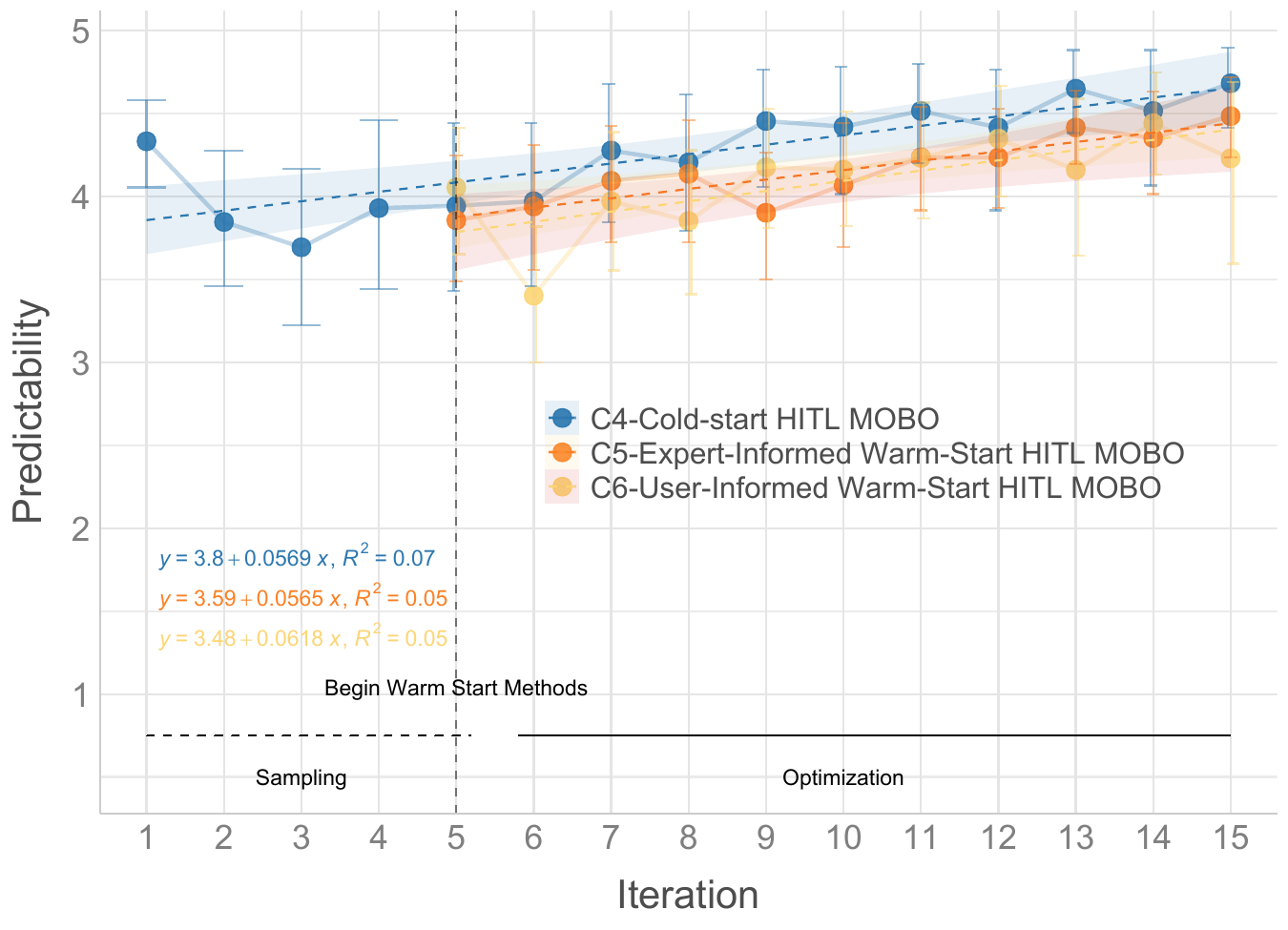}
   \caption{Progression of \textbf{predictability} values over the MOBO iterations.}
   \label{fig:runs_pred}
    \Description{}
  \end{subfigure}
    \caption{Value progression of trust and predictability.}
    \Description{Progression of trust (going steadily upward) and predictability (going steadily upward) over the MOBO iterations.}
    \label{fig:run_trust_pred}
\end{figure*}

\begin{figure*}[ht!]
\centering
             \begin{subfigure}[b]{0.49\linewidth}
             \includegraphics[width=\linewidth]{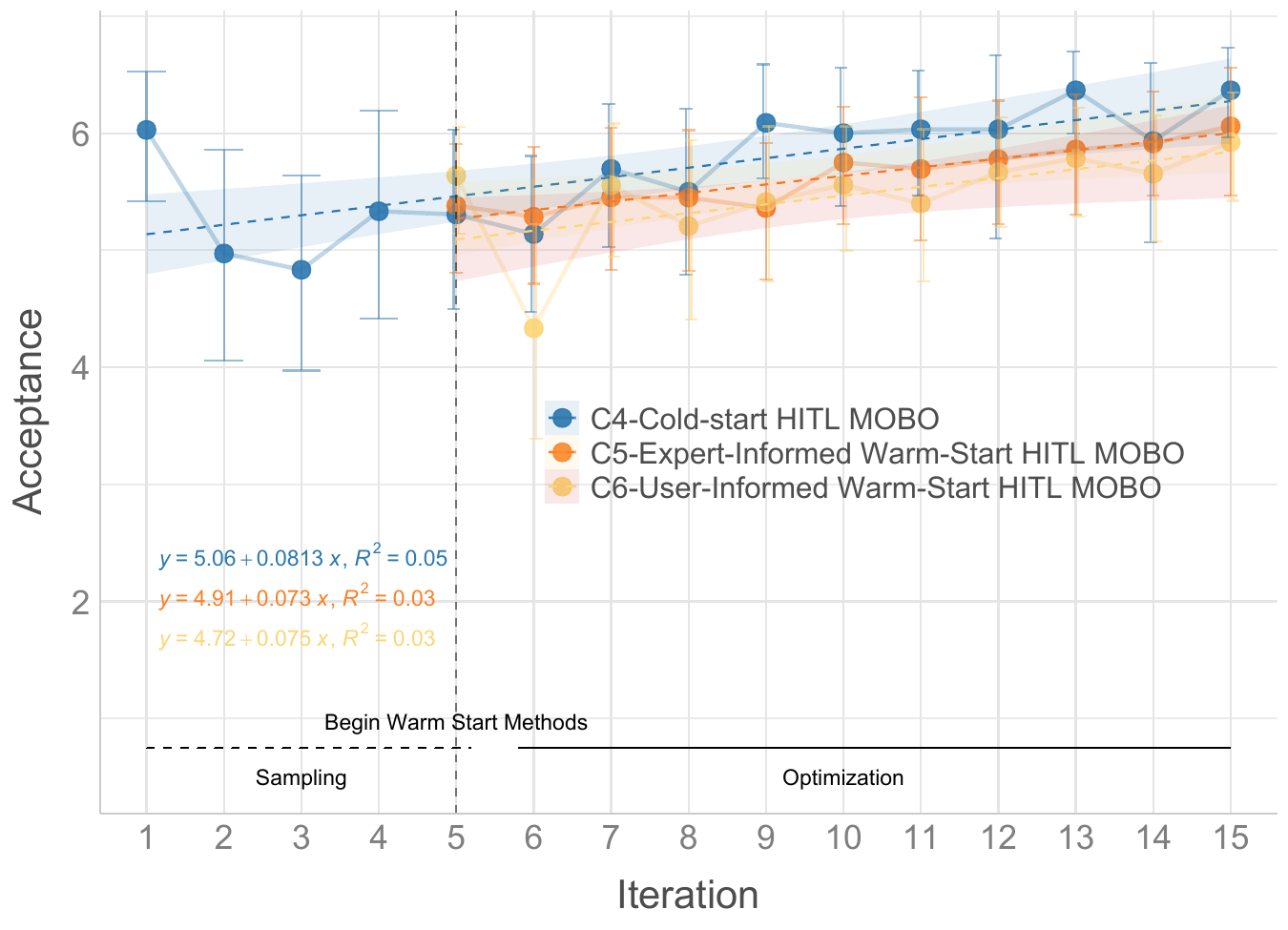}
   \caption{Progression of \textbf{acceptance} values over the MOBO iterations.}
   \label{fig:runs_acc}
    \Description{Progression of acceptance (going steadily upward) over the MOBO iterations.}
              \end{subfigure}
         \begin{subfigure}[b]{0.49\linewidth}
    \includegraphics[width=\linewidth]{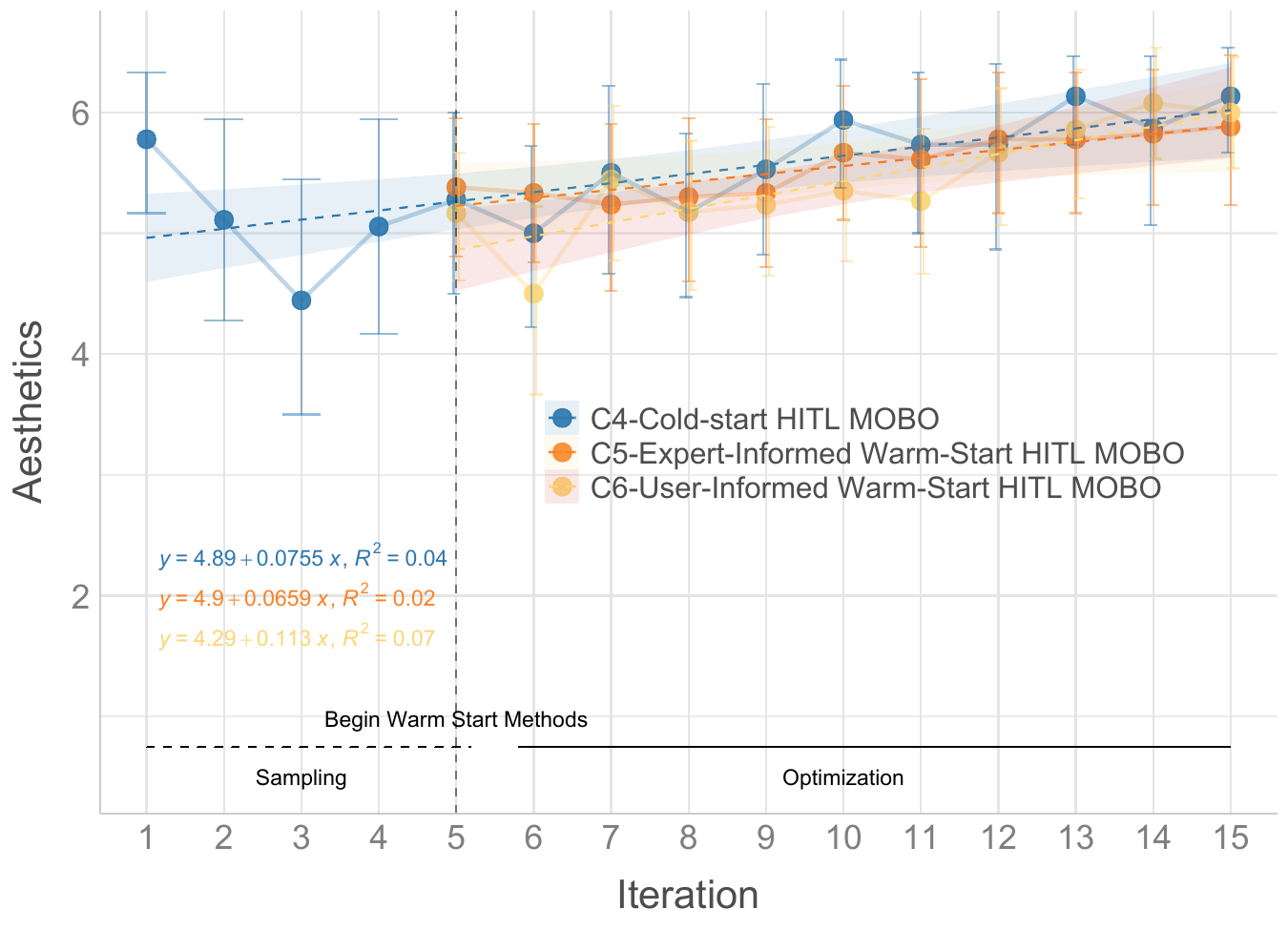}
   \caption{Progression of \textbf{aesthetics} values over the MOBO iterations.}
   \label{fig:runs_aes}
    \Description{Progression of aesthetics (going steadily upward) over the MOBO iterations.}
  \end{subfigure}
    \caption{Value progression of acceptance and aesthetics.}
    \Description{Progression of acceptance (going steadily upward) and aesthetics (going steadily upward) over the MOBO iterations.}
    \label{fig:run_acc_aes}
\end{figure*}


\subsection{Eye-Tracking Results}

\begin{figure*}[ht!]
\centering
    \includegraphics[width=.75\linewidth]{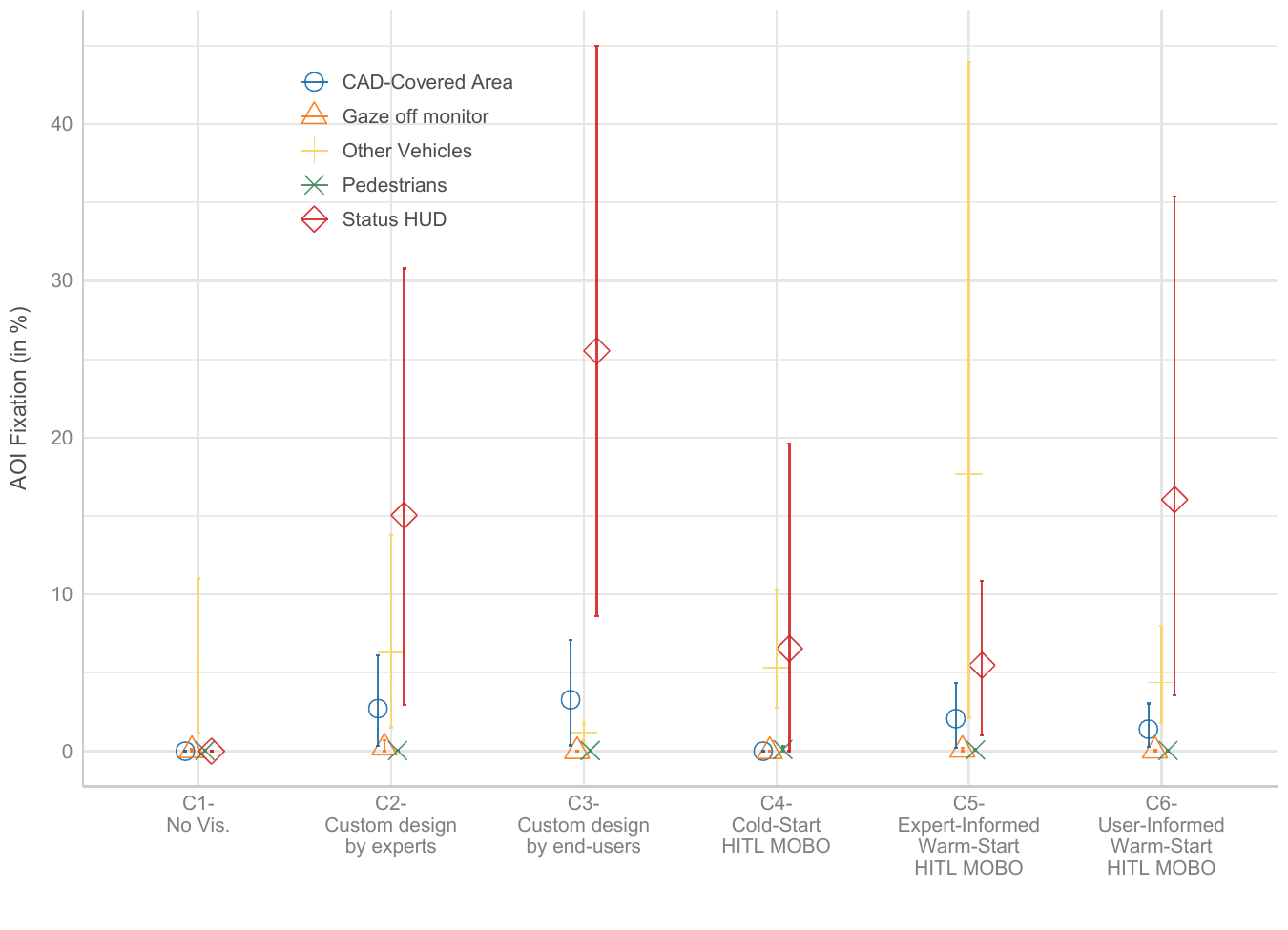}
   \caption{Eye fixations without the instances that the participants looked at the monitor but not at an AOI. For C6-User-Informed Warm-Start HITL MOBO, C2-Custom design by experts, and C3-Custom design by end-users, particular emphasis was placed on the speedometer. In C5-Expert-Informed Warm-Start HITL MOBO, emphasis was put on the car, which was also gazed upon comparatively frequently in the other conditions. }
   \label{fig:eye_gaze}
    \Description{This figure shows the overview of the fixations (in percent) of the areas of interest car, CAD-covered area, offscreen, people, and speedometer. }
\end{figure*}

\end{document}